%% file: Planetesimal_pebble.tex
\documentclass[printer]{aa}
\usepackage{graphicx,natbib}
\usepackage{tikz}
\usetikzlibrary{positioning}
\usepackage{amsmath}
\usepackage[latin1]{inputenc}
\usepackage[T1]{fontenc}
\bibpunct{(}{)}{;}{a}{}{,}
\usepackage{txfonts}

\include{def}
\usepackage{bm}
\usepackage{units}
\usepackage{color}
\graphicspath{{./fig/}}
\usepackage{adjustbox}
\usepackage{longtable}

\begin{document}

\title{Pebbles versus planetesimals: \\
the outcomes of population synthesis models}

\author{N. Br\"ugger$^1$, R. Burn$^1$, G.A.L. Coleman$^{1,2}$, Y. Alibert$^1$, W. Benz$^1$}
\offprints{N. Br\"ugger}
\institute{1. Physikalisches Institut, Universit\"at Bern, CH-3012 Bern, Switzerland \\
2. Astronomy Unit, Queen Mary University of London, Mile End Road, London, E1 4NS, U.K.\\
\email{natacha.bruegger@space.unibe.ch }
}

\abstract
{In the core accretion scenario of giant planet formation, a massive core forms first and then accretes a gaseous envelope. 
When discussing how this core forms some divergences appear. First scenarios of planet formation predict the accretion of km-sized bodies, called planetesimals, while more recent works suggest growth by accretion of pebbles, which are cm-sized objects.} 
{These two accretion models are often discussed separately and we aim here at comparing the outcomes of the two models with identical initial conditions.} 
{The comparison is done using two distinct codes: one computing the planetesimal accretion, the other one the pebble accretion. All the other components of the simulated planet growth are computed identically in the two models: the disc, the accretion of gas and the migration. Using a population synthesis approach, we compare planet simulations and study the impact of the two solid accretion models, focussing on the formation of single planets.} 
{We find that the outcomes of the populations are strongly influenced by the accretion model. The planetesimal model predicts the formation of more giant planets, while the pebble accretion model forms more super-Earth mass planets.
This is due to the pebble isolation mass ($\rm M_{\rm iso}$) concept, which prevents planets formed by pebble accretion to accrete gas efficiently before reaching $\rm M_{\rm iso}$.
This translates into a population of planets that are not heavy enough to accrete a consequent envelope but that are in a mass range where type I migration is very efficient.
 We also find higher gas mass fractions for a given core mass for the pebble model compared to the planetesimal one caused by luminosity differences. This also implies planets with lower densities which could be confirmed observationally.}
{We conclude that the two models produce different outputs. Focusing on giant planets, the sensitivity of their formation differs: for the pebble accretion model, the time at which the embryos are formed, as well as the period over which solids are accreted strongly impact the results, while the population of giant planets formed by planetesimal accretion depends on the planetesimal size and on the splitting in the amount of solids available to form planetesimals.}

\titlerunning{Planetesimal vs pebble accretion}
\authorrunning{Br\"ugger et al.}
\maketitle

\section{Introduction}
\label{introduction}

In the standard giant planet formation theory, the so-called core-accretion model, a core forms first through the accretion of solids and then, if it becomes massive enough, it accretes gas.
A crucial constraint for gas accretion is that the core should be massive enough to accrete the gas before the dissipation of the gas disc \citep{Haisch01}.
The first scenarios predict that the solids accreted by the core are planetesimals, which are $\sim$kilometer-sized objects \citep{Pollack, Fortier13}.
Historically the typical radius of planetesimals was 100 km.
One problem that arises when using planetesimals of this size, is that the time needed to form a core is typically longer than expected disc lifetimes \citep{Pollack}.
Forming giant planets is therefore difficult for traditional planet formation models \citep{ColemanNelson14}.
Reducing the size of the planetesimals allows however to form cores within typical disc lifetimes \citep{ColemanNelson16,ColemanNelson16b}.
This time-scale struggle gave birth to a new approach that suggests the accretion of drifting centimeter-sized bodies known as pebbles \citep{Birnstiel12}.
Due to their small size, pebbles are able to be accreted much more easily through increased gas drag, resulting in a more rapid core formation \citep{OrmelKlahr2010,Lambrechts12}.

These two scenarios of solid accretion were recently compared by \citet{Coleman19} with the aim of examining planet formation around low mass stars akin to the Trappist-1 planetary system.
They explored a wide range of initial conditions and found that both scenarios formed remarkably similar planetary systems, in terms of planetary masses and periods, resonances between neighbouring planets, and the general observability of the planets and their respective systems.
Whilst \citet{Coleman19} compared the two scenarios within the frame of the Trappist-1 system, in this paper we focus on solar mass stars and vary some parameters of our model, e.g. the starting time of the embryo or the distribution of the amount of solids.
We aim here at comparing the two solid accretion scenarios by using identical initial conditions drawn form a distribution comparable to those used within population synthesis models \citep[e.g. see][]{Mordasini15}.
Using two separate models, one for planetesimal accretion and one for pebble accretion, we examine the outcomes of population of single planet systems.
To proceed, we use the same disc model, gas accretion model and migration regimes for both codes.
It is important to underline that the two codes are distinct from one another and that this comparison aims at comparing the outcomes of the two different accretion scenarios and not to achieve a match to observations.

This paper is organised as follows.
In Sect. \ref{theoreticalmodel} we provide all of the theoretical aspects behind the comparison. We discuss the similarities between the two codes, e.g. the disc model and its evolution, the gas accretion theory and the migration formulae, as well as the two distinct accretion models.
To test our implementations we present in Sect. \ref{comparisonbetweenthemodels} comparisons between the two codes for the common components of the models. 
The evolution of the disc is discussed, as well as the accretion of gas and the migration regimes.
Once the agreement between the two codes is established, the effect of the two solid accretion models can then be compared.
Using a population synthesis approach, we compute single planet per disc simulations and study the outcomes in Sect. \ref{populationsynthesisoutcomes}, where we also compare the two modes of solid accretion.
Finally, Sect. \ref{conclusion} is dedicated to discussions and conclusions.


\section{Theoretical models}
\label{theoreticalmodel}

We first introduce the disc model, which is common to both accretion models.
We then present the planetesimal accretion model, which is an improved version of that presented in \cite{Mord12a,Alibert13,Fortier13}, as well as the pebble accretion model, which is similar to that of \citet{Brugger18}.
We then describe another common aspect of the two models: the gas accretion.
Finally, we discuss the planet migration.

\subsection{Disc model and evolution}
\label{discmodel_theory}

The disc model we use is similar to that provided by \citet{Hueso05}.
The initial gas surface density profile follows \citep{Andrews10}:
\begin{equation}
\Sigma(r) = \Sigma_0 \left( \frac{r}{5.2 \textrm{ AU}} \right)^{-\beta} \exp \left[ - \left( \frac{r}{r_{\textrm{out}}} \right)^{(2-\beta)} \right] \, ,
\end{equation}

where $\Sigma_0$ is the initial surface density at 5.2 AU and $r$ is the location in the disc, $r_{\rm out}$ is the outer radius of the disc and $\beta = 0.9$.
This disc model accounts for observational constraints that are relevant to the disc evolution calculations (stellar properties, disc outer radius and surface density profile or accretion rate).
The disc profile is therefore very different from that provided by \citet{Bitsch15a} and used in \citet{Brugger18}, which may lead to different outcomes.
For instance the surface density in the outer regions is much lower in the disc used here compared to that  of \citet{Bitsch15a} .

To calculate the midplane temperature we use a one-dimensional model based on a semi-analytical approach, where we include both stellar irradiation and the dissipation of viscous energy for heating the disc.
In the radial direction the disc is assumed to be thick.
Heat can therefore be more efficiently transported vertically where the disc can be geometrically thin or thick.
Consequently these two regimes are both combined in the midplane temperature $\rm T_{m}$ determination \citep{Nakamoto94,Hueso05}:
\begin{equation}
T_m^4 =  \frac{1}{2 \sigma} \left( \frac{3 \kappa_R}{8} \Sigma  + \frac{1}{2 \kappa_p \Sigma} \right) \dot{E}_{\nu}  + T_{\rm irr}^4 \, ,
\label{Centraltemperature}
\end{equation}
with $\sigma$ being the Stefan-Boltzmann constant, $\kappa_R$ the Rosseland mean opacity, $\kappa_P$ the Planck opacity, $\Sigma$ the gas surface density of the disc, $\dot{E}_{\nu} =  \frac{9}{4} \Sigma \nu \Omega_K^2$ the viscous energy dissipation rate \citep{Nakamoto94} and $T_{\rm irr}$ the effective temperature due to stellar irradiation that is a function of the stellar temperature $T_{*} $ \citep{Adams88,Ruden91,Hueso05}:
\begin{equation}
 T_{\rm irr} = T_{*} \left( \frac{2}{3 \pi} \left(\frac{R_*}{r} \right)^3 + \frac{1}{2} \left( \frac{R_*}{r} \right)^2 \frac{H}{r} \left(\frac{\rm dln(\mathit{H})}{\rm dln(\mathit{r})} -1 \right) \right)^{1/4} \, .
 \label{Irradiationtemperature}
\end{equation}
Here $\Omega_K = \sqrt{\frac{GM_*}{r^3}}$ is the Keplerian frequency, $T_{*}$ is the star's temperature, $R_{*}$ is the radius of the star (see Table \ref{tabledisc}), $H$ the disc scale height and $\frac{\rm dln(\mathit{H})}{\rm dln(\mathit{r})} = \frac{9}{7}$, which is the equilibrium solution for a disc where the flaring term (term containing $\frac{\rm dln(\mathit{H})}{\rm dln(\mathit{r})}$ in the temperature determination (Eq. \ref{Irradiationtemperature})) is the dominant one \citep{Hueso05}.
The vertical structure of the disc can then be derived from Eq. \ref{Centraltemperature}, the viscosity $\nu$ and the opacity of the disc $\kappa$ \citep{Bell94}, which in our model is scaled with the amount of dust in the disc.

Once the properties of the disc are defined, its evolution follows the standard diffusion equation \citep{Lynden-BellPringle1974}:
\begin{equation}
\frac{\partial \Sigma}{\partial t} = \frac{1}{r} \frac{\partial}{\partial r}  \left[ 3r^{1/2} \frac{\partial}{\partial r} (\nu \Sigma r^{1/2}) \right] \, ,
\label{evolutionsurfacedensitydisc}
\end{equation}
where $\nu=\alpha c_s H$ is the viscosity, which is parametrized using the $\alpha$-viscosity parameter (chosen to be $\alpha = 0.002$) of \citet{Shak} and the isothermal sound speed $c_s$. 

To obtain realistic disc lifetimes (between 2 and 5 Myr \citep{Haisch01}), we use the external photoevaporation model of \citet{Matsuyama03} and the internal photoevaporation model given by \citet{Clarke2001} with modifications from \citet{AlexanderPascucci12}.	 
For internal photoevaporation, \citet{Clarke2001} assume a region within which the photoionized gas remains bound to the star. 
This region is defined by its radius:
\begin{equation}
\label{R}
R_{g, \rm int} = \frac{GM_*}{c_s} \, ,
\end{equation}
with $c_s$ being the sound speed of photoionized gas ($\rm T = 1000$ K) and $\rm M_*$ the mass of the star (see Table \ref{tabledisc}).
Beyond this radius, material can be lost from the disc at a rate given by \citep{Clarke2001}:
\begin{equation}
\dot{\Sigma}_{w, \rm int} = 2 c_s n_0(r) m_H \, ,
\end{equation}
where the factor $2$ considers the mass loss from both sides of the disc, $n_0(r)$ is the number density at a distance $r$ and $m_H$ is the mass of the hydrogen atom.
This corresponds to a total wind mass-loss rate of \citep{Clarke2001}:
\begin{equation}
\dot{M}_{w, \rm int} = 4.1 \times 10^{-10} \phi_{41}^{1/2} \left( \frac{M_{*}}{M_{\odot}} \right)^{1/2} \rm M_{\odot} \rm yr^{-1} \, ,
\end{equation}
where $\phi_{41} = 1$ is the ionizing photon flux of the star in units of $10^{41} \rm s^{-1}$.\\
For external photoevaporation, \citet{Matsuyama03} predicts that the surface density evaporation rate for radii beyond $R_{g, \rm ext}$ (same definition as Eq. \ref{R} but with a sound speed given for a temperature of $T=10^4$ K) follows:
\begin{equation}
\dot{\Sigma}_{w, \rm ext} = \frac{\dot{M}_{w, \rm ext}}{\pi \left( R_d^2 - \beta^2 R_g^2 \right)} \, ,
\end{equation}
where $R_d$ is the disc outer edge ($R_d = 1000$ AU in our test cases, see Table \ref{tabledisc}), $\beta = R_*/R_g$ and the mass loss rate is given by $\dot{M}_{w, \rm ext} = 1 \times 10^{-7}$ $\rm M_{\odot}/ \rm year$ for our test cases (see Table \ref{tabledisc} as well).

Regarding the solid components of the disc, the total amount of solids available in the disc $Z_{\rm tot}$, initially all in the form of dust, is split into a fraction that forms the bodies that can be accreted (either planetesimals or pebbles) while the rest remains as dust, contributing to the disc opacity.
The same splitting is applied in both models and the two ratios we investigate are  $Z_{\rm peb, \rm plan} = 0.9 \times Z_{\rm tot}$ with $Z_{\rm dust} = 0.1 \times Z_{\rm tot}$, which we call the $\varepsilon = 0.9$ case, and $Z_{\rm peb, \rm plan} = 0.5 \times Z_{\rm tot}$ with $Z_{\rm dust} = 0.5 \times Z_{\rm tot}$, which we call the $\varepsilon = 0.5$ case.
For our test cases (Sect. \ref{comparisonbetweenthemodels}), we use the $\varepsilon = 0.9$ case following \citet{Brugger18} and the total fraction of solids to gas is given by $Z_{\rm tot} = 0.01$ (see Table \ref{tabledisc}).

Another component that is common to both models is the determination of the ice line. For simplicity we define it as the place in the disc where the temperature is equal to 170 K \citep{Burn19}. 
This location therefore depends on the temperature of the disc, which is influenced by the opacity of the disc. 
The latter is impacted by the amount of solids as well as the ratio $\varepsilon$, since the fraction that remains as dust contributes to the disc opacity.
The ice line location has an impact on the pebble size (see Sect. \ref{pebbleaccretionmodel}) and on the composition of the planets (see Sect. \ref{Distribution}).

\subsection{Planetesimal accretion model}
\label{planetesimalaccretionmodel}
The planetesimal accretion model is described in detail in \citet{Fortier13}.
 The basic principle is to represent planetesimals as a fluid-like disc.
The initial profile of the surface density of planetesimal $\Sigma_{\rm pls}$ is however steeper than the one of the gas \citep{Lenz19,Drazkowska17}.
The surface density as well as the eccentricity rms $e_{\rm pls}$ and the inclination rms $i_{\rm pls}$ evolve over time. To have a consistent description of $e_{\rm pls}$ and $i_{\rm pls}$ for all planetesimal sizes, we solve the differential equations for self-stirring \citep[e.g.][]{Ohtsuki99}, the gravitational stirring of planetesimals by forming planets \citep{Ohtsuki99} as well as the damping by gas drag \citep{Adachi,Inaba01,Rafikov04} instead of assuming that equilibrium between stirring and damping is attained instantaneously.\footnote{\citet{Fortier13} found that for larger planetesimal sizes (10 km or 100 km), the assumption of equilibrium $e_{\rm pls}$ and $i_{\rm pls}$ is justified, but here we assume smaller planetesimal sizes (1 km).} 
We do not take into account the radial drift of planetesimals as it was found to be negligible over the disc lifetime for our chosen radius of 1 km. 
This approach is valid for particles that decouple from the gas, which typically happens at sizes larger than \unit{100} {m} \citep{Burn19}. 

The accretion of solids is given by 
\begin{equation}
\dot{M}_{\rm pls} = \Omega_K \bar{\Sigma}_{\rm pls} R_H^2 p_\text{coll}\, ,
\end{equation}
where $\Omega_K$ is the Keplerian angular velocity, $R_H = \left( \frac{m_p + m_{\rm pls}}{3M_*} \right)^{1/3} a$ is the planet's Hill Radius, $\bar{\Sigma}_{\rm pls}$ is averaged over the planet's feeding zone (spanning ten Hill radii for a planet on a circular orbit, considering that the planet is in the middle of its feeding zone) of the aforementioned surface density of planetesimals and $p_\text{coll}$ is the collision probability following
\citet{Inaba01}:
\begin{equation}
\label{eq:P_col}
p_\text{coll}=\min\left( {p}_\text{med},\left({p}^{-2}_\text{high}+{p}^{-2}_\text{low}\right)^{-1/2}\right)\,.
\end{equation}
The individual components are:
\begin{equation}
\label{eq:P_col_high}
p_\text{high} = \frac{\tilde{r}_p^2}{2\pi} \left(\mathcal{F}(I)+\frac{6}{\tilde{r}_p}\frac{\mathcal{G}(I)}{({{\tilde{e}}})^2}\right)\,,
\end{equation}
\begin{equation}
p_\text{med} = \frac{\tilde{r}_p^2}{4\pi \tilde{i}} \left(17.3+\frac{232}{\tilde{r}_p}\right) \, ,
\end{equation}
\begin{equation}
p_\text{low} = 11.3\sqrt{\tilde{r}_p}\,.
\end{equation}

Here $I \equiv i_{\rm pls}/e_{\rm pls}$, $\tilde{e} = \frac{a ^. e_{\rm pls}}{R_H}$ is the eccentricity of the planetesimals in Hill's unit, we use numerical fits for the integrals $\mathcal{F}(I)$ and $\mathcal{G}(I)$ following \citet{Chambers06}, $\tilde{i} = \frac{a ^. i_{\rm pls}}{R_H}$ is the inclination of the planetesimals in Hill's unit and 
\begin{equation}
\tilde{r}_p \equiv \frac{R_\text{capture}+R_{\rm pls}}{R_H}\,.
\end{equation}
$R_{\rm pls} = 1$ km is the planetesimal radius and $R_\text{capture}$ is the planet's capture radius, which is enlarged as described in \citet{Inaba} when a gaseous envelope is present. 
We numerically retrieve $R_\text{capture}$ from equation (17) of \citet{Inaba}:
\begin{equation}
R_{\rm pls} = \frac{3}{2} \frac{v_\infty^2 + 2 G M_\text{core}/R_\text{capture}}{v_\infty^2 + 2 G M_\text{core}/R_H} \frac{\rho(R_\text{capture})}{\rho_{\rm pls}}\, .
\end{equation}
Here $\rho_{\rm pls}$ is the density of the planetesimal, $\rho(R_\text{capture})$ is the density of the gaseous planetary envelope at $R_\text{capture}$ and 
\begin{equation}
v_\infty = v_{K} \sqrt{5/8\, e_{\rm pls}^2 + i_{\rm pls}^2}
\end{equation}
is the typical relative velocity at infinite distance to the planet.
The Keplerian velocity $v_{K}$ is defined as $v_{K} = \sqrt{\frac{GM}{R}}$.

In addition to the accreted mass of planetesimals that is reduced from $\Sigma_{\rm pls}$ over the planet's feeding zone, an estimated amount of ejected planetesimals is subtracted following \cite{IdaLin2004}
\begin{equation}
      \dot{M}_{\text{ejected,} \rm pls} = \left(\frac{a_\text{planet} M_\text{planet} }{2 M_* R_\text{capture} }\right)^2 \dot{M}_{\rm pls}\,.
\end{equation}
The factor in front of the planetesimal accretion rate is the ratio of the characteristic surface speed and the escape speed from the star.

\subsection{Pebble accretion model}
\label{pebbleaccretionmodel}

For the pebble accretion model we follow the model outlined by \citet{Brugger18}. 
An embryo is assumed to form via the streaming instability in the disc at a given time, which is a free parameter of the model.
This embryo grows by accreting pebbles that form in the outer regions of the disc and then drift towards the star \citep{Lambrechts14}. 
The amount of pebbles depends on the fraction of solids in the disc that can turn into pebbles ($Z_{\rm peb}$) as mentioned in Sect. \ref{discmodel_theory}.

We use the pebble accretion rates given by \citet{Johansen17} which distinguish between the Bondi accretion regime (small protoplanets) and the Hill accretion regime (large protoplanets).
The Bondi accretion regime occurs for low mass planets where the planets do not accrete all of the pebbles that pass through their Hill sphere, i.e. the planet's Bondi radius is smaller than the Hill radius.
Once the Bondi radius becomes comparable to the Hill radius, the accretion rate becomes Hill sphere limited, and so the planet accretes in the Hill accretion regime. This is the typical regime for more massive bodies in the disc.
Within the Hill regime a further distinction occurs whether the planet is accreting in a 2D or a 3D mode.
This is dependent on the relation between the Hill radius of the planet and the scale height of the pebbles in the disc.
For planets with a Hill radius smaller than the scale height of pebbles, the accretion is in the 3D mode, whilst for planets with a Hill radius larger than the pebble scale height, the 2D mode.
The general equation for the 2D and 3D accretion rates are respectively \citep{Johansen17}
\begin{equation}
\dot{M}_{\rm 2D} = 2 R_{\rm acc} \Sigma_{\rm peb} \delta v \, ,
\label{Mdotpebbles_2D}
\end{equation}
and:
\begin{equation}
\dot{M}_{\rm 3D} = \pi R_{\rm acc}^2 \rho_{\rm peb} \delta v \,,
\label{Mdotpebbles_3D}
\end{equation}
where $\rho_{\rm peb}$ is the midplane pebble density and $\Sigma_{\rm peb} = \frac{\dot{M}_{\rm peb}}{2 \pi R v_r}$ is the pebble surface density including the flux of pebbles $\dot{M}_{\rm peb}$ and their velocity $v_r$. 
The approach speed is given by $\delta v = \Delta v + \Omega_K R_{\rm acc}$, with $\Delta v \sim \eta v_K$ being the sub-Keplerian velocity, $\eta = - \frac{1}{2} \left(\frac{H}{r} \right)^2 \frac{\rm dln \mathit{P}}{\rm dln \mathit{r}}$ the gas pressure gradient and $\Omega_K$ the Keplerian frequency. 
The accretion radius $R_{\rm acc}$ used in Eqs. \ref{Mdotpebbles_2D} and \ref{Mdotpebbles_3D} is defined with the help of:
\begin{equation}
 R_{\rm acc}' = \left( \frac{4 \tau_f}{t_B} \right)^{1/2} R_B \,,
 \label{Racc_Bondi}
\end{equation}
in the Bondi regime, and:
\begin{equation}
 R_{\rm acc}' = \left(  \frac{\Omega_K \tau_f}{0.1} \right)^{1/3} R_H \,,
 \label{Racc_Hill}
\end{equation}
in the Hill regime.

Here $R_B = \frac{GM}{\Delta v^2}$ is the Bondi radius and $t_B = R_B/ \Delta v$. $R_H$ is the Hill radius and $\tau_f = \rm St/ \Omega_K$ \citep{Johansen17} with $\rm St$ being the Stokes number that describes the pebble size (\citealp{Lambrechts14}, see discussion below).
These expressions (Eq. \ref{Racc_Bondi} and \ref{Racc_Hill}) however only consider strong coupling between the pebbles and the protoplanet.
In order to account for the less efficient accretion when the friction time becomes longer than the time to drift past the protoplanet, $R_{\rm acc}$ becomes \citep{OrmelKlahr2010}:
\begin{equation}
R_{\rm acc} = R_{\rm acc}' \rm e^{ -0.4 (\tau_f / t_p )^{0.65}},
\end{equation}
before going back to Eqs. \ref{Mdotpebbles_2D} and \ref{Mdotpebbles_3D}. Here $t_p = GM/(\Delta v + \Omega_K R_H)^3$ is the drifting time-scale. 

The pebble size is usually described by the Stokes number $\rm St$.
Outside the ice line the pebbles are assumed to be made of ice surrounding trapped silicates. Their size is given by $t_{\rm growth} (r_g) = t_{\rm drift}(r_g)$, leading to $\rm St \sim 0.01 - 0.1$ \citep{Lambrechts14}.
However inside the ice line, this assumption no longer holds because the ice sublimates \citep{IdaGuillot16} and releases the silicates. 
Therefore the pebble size significantly shrinks to the size of these silicate grains, which are much smaller than the original icy pebbles \citep{Morbi15,Shibaike19}.
Observations hint that the size of these silicates is similar to the one of chondrules, which are mm-sized particles \citep{Friedrich2015}.
Therefore if a planet accretes pebbles inside the ice line, the accreted pebbles have a much lower Stokes number $\rm St << 1$ \citep{Birnstiel12}, which impacts on the accretion rate (see discussion in Sect. \ref{Massvssemimajoraxis}).

The embryo thus grows by accreting pebbles until it reaches the so-called pebble isolation mass \citep{Lambrechts14} (see also \citealp{Ataiee2018,Bitsch18}):
\begin{equation}
M_{\rm iso} = 20 \left( \frac{H/R}{0.05} \right)^3 \cdot M_{\oplus} \,.
\end{equation}
The pebble isolation mass is the mass required to perturb the gas pressure gradient in the disc.
Thus the gas velocity becomes super-Keplerian in a narrow ring outside the planet's orbit reversing the action of the gas drag.
The pebbles are therefore pushed outwards rather than inwards and accumulate at the outer edge of this ring stopping the core from accreting solids \citep{PaardekooperMellema06}.
Consequently the planet begins to accrete gas more efficiently.
Therefore the calculation of the envelope structure (presented in Sect. \ref{gasaccretionmodel_theory}) starts at the $\rm min(M_{\rm iso}, 3 M_\oplus)$.

\subsection{Gas accretion model}
\label{gasaccretionmodel_theory}

The computation of gas accretion is similar in both planetesimal and pebble models. 
The internal structure of the planetary envelope is computed by solving the following equations :
\begin{equation}
\frac{\partial m}{\partial r}= 4 \pi r^2 \rho \,,
\label{massconservation}
\end{equation}
\begin{equation}
\frac{\partial P}{\partial r} = - \frac{Gm}{r^2} \rho \,,
\label{evolutionofpressurewithdepth}
\end{equation}
and:
\begin{equation}
\frac{\partial T}{\partial r}= \frac{T}{P} \frac{dP}{dr} \nabla \,,
\label{energytransfer}
\end{equation}
which represent the mass conservation, the equation of hydrostatic-equilibrium and energy transfer respectively \citep{BodenheimerPollack86,Alibert05,Mord12a,Alibert16,CPN17}.
The pressure $P$ and temperature $T$ depend on the mass $m$ included in a sphere of radius $r$. 
The density $\rho(P,T)$ follows \citet{Saumon95} and the temperature gradient depends on the stability of the zone against convection: for convective zones, it is assumed to be given by the adiabatic gradient.
Therefore $\nabla = \frac{d \rm ln(T)}{d \rm ln(P)} = \rm min(\nabla_{\rm ad}, \nabla_{\rm rad})$ where
\begin{equation}
\nabla_{\rm ad} = \frac{\rm dln(\mathit{T})}{\rm dln(\mathit{P})} \, ,
\end{equation}
\begin{equation}
\nabla_{\rm rad}  = \frac{3}{64 \pi \sigma G} \frac{\kappa L P}{T^4 m} \,,
\label{nablarad}
\end{equation}
with $\kappa$ \citep{Bell94} being the full interstellar opacity (see however Sect. \ref{populationsynthesisoutcomes}) and $L$ being the luminosity of the planet computed by energy conservation and including the solid accretion luminosity, the gas contraction luminosity and the gas accretion luminosity \citep{Mord12a,Mord12b,Alibert13}.

The mass of the envelope is then determined by iteration. Comparing the envelope masses between two iterations provides the gas accretion rate \citep{Alibert05}.
For runaway gas accretion \citep{Pollack}, the maximum accretion rate is limited by what can be provided by the disc:
\begin{equation}
\dot{M}_{\rm gas, \rm max} = \dot{M}_{\rm disc} = 3 \pi \nu \Sigma \,.
\end{equation}

\subsection{Planet migration}
\label{planetmigration_theory}

As planets grow, they interact gravitationally with the surrounding gas, exchange angular momentum and migrate through the disc. 
Low-mass planets that are embedded in the disc feel a torque arising from the gravitational interaction between the planet and the disc.
This process is called type I migration.
The torque felt by the planets is the composition of the Lindblad torque $\Gamma_L$ and the corotation torque $\Gamma_c$ \citep{pdk10,pdk11}
\begin{equation}
\label{Gamma_tot}
\Gamma_{\rm tot} = \Gamma_L + \Gamma_c \,.
\end{equation}
The Lindblad torque is a torque exerted by density waves on the planet.
The presence of the planet creates these waves in the disc at locations called Lindblad resonances.
On the other hand the corotation torque corresponds to an exchange of angular momentum between the planet and the neighbouring gas situated in the corotation region of the planet.
The two torques depend on the local gradients of surface density, temperature and entropy. 
In locations where a strong negative temperature gradient is present, the planet is expected to migrate outwards. These regions of outward migration lie where $|\Gamma_c| > |\Gamma_L|$. \\

Higher mass planets on the other hand are able to open a gap in the disc \citep{LinPapaloizou86}. 
This slows down their migration towards the star.
The gap opening depends on the scale height and viscosity of the disc.
A gap opening criterion is provided by \cite{Crida}:
\begin{equation}
P = \frac{3}{4} \frac{H}{r_H} + \frac{50}{qRe} \leq 1 \,,
\end{equation}
where $q = M_p/M_\star$ is the mass ratio and $Re$ is the Reynolds number given by $Re = r_p^2 \Omega_K^2 / \nu$. 
If the planet fulfils this criterion, it starts to migrate towards the star in the so-called type II migration regime on a time-scale that is a function of the viscosity of the disc $\nu$ \citep{Mords09}:
\begin{equation}
\label{tauII}
\tau_{II} = \frac{2a_p^2}{3 \nu} \times \rm max \left( 1, \frac{M_p}{2 \Sigma_{\rm gas} a_p^2} \right) \,.
\end{equation} 
The maximum term allows the so-called planet dominated regime to be taken into account.
This regime is a consequence of the decrease in the gas disc mass and the slowing down of migration as the planet becomes more massive.

\subsection{Long-term evolution}
\label{Longtermevolution}

Once the gas disc has disappeared, the planets enter the evolution stage.
At this point both gas accretion and disc-driven migration cease.
We take the outcomes of our populations as initial conditions for this long-term evolution.
Our aim is to obtain the density of the planets.
To get realistic radii in addition to the known masses we use the evolution model of \citet{Mord12b,Mord12a} including atmospheric loss due to photoevaporation \citep{JinMordasini14}.
The outer radius of the numerical envelope structure extends to very low densities. 
Therefore, we follow the prescription of \citet{Hansen2008} to calculate what radius would be observed by a generic transit observation.


\section{Comparisons between the models}
\label{comparisonbetweenthemodels}

In order to perform a proper comparison between the two separate models of solid accretion, all the other components of the simulated planet growth should be similar: e.g. the disc model, the accretion of gas and the migration of the planet.
Therefore we complete tests to consolidate both models and make sure that they are identical in these aspects.

\begin{center}
\begin{table}
\centering
\caption{System properties used in all test cases.}
\label{tabledisc}
\begin{tabular}{|c|c|}
\hline
\textbf{System properties} & \textbf{Values}  \\
\hline
Disc mass & 0.017 $\rm M_{\odot}$ \\
\hline
Slope & 0.9  \\
\hline
$\alpha$ & 0.002 \\
\hline
$\rm Z_{\rm tot}$ & 0.01  \\
\hline
$\mu$ & 2.27  \\
\hline
Inner edge of the disc & 0.1 AU  \\
\hline
Outer radius of the disc $\rm R_d$ & 1000 AU  \\
\hline
Cut off radius of the disc & 30 AU \\
\hline
Photo-evaporation rate $\dot{M}_{w, \rm ext}$ & $1 \times 10^{-7}$ $\rm M_{\odot} / \rm year$ \\
\hline
$\rm R_*$ & 2 $\rm R_\odot$  \\
\hline
$\rm M_*$ & 1 $\rm M_\odot$  \\
\hline
$\rm T_*$ & 4480 K  \\
\hline
\end{tabular}
\end{table}
\end{center}

\subsection{Disc model}
\label{discmodels}

\begin{figure}
\hspace{0cm} \includegraphics[height=0.35\textheight,angle=0,width=0.35\textheight]{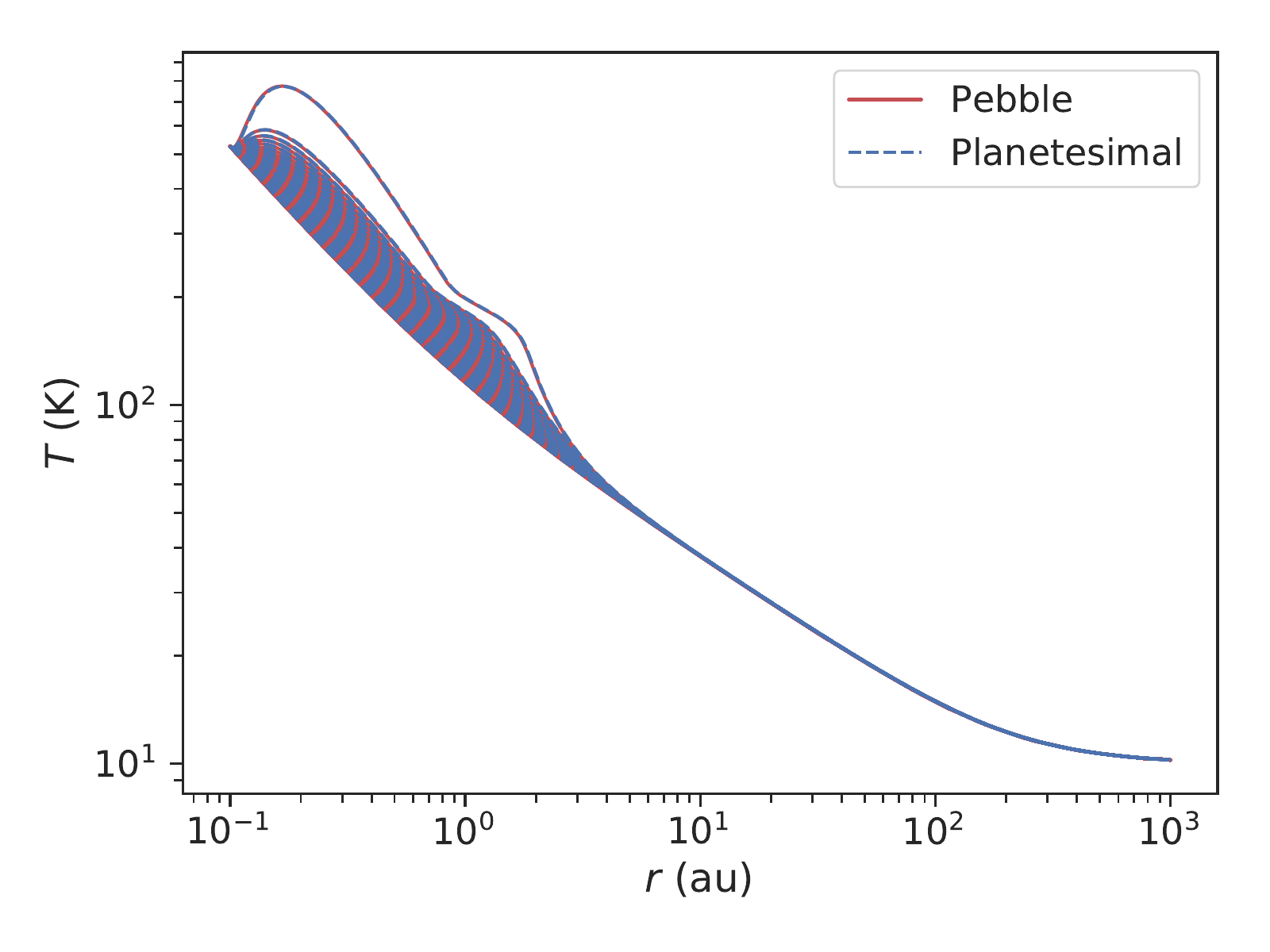}
\caption{Temperature profile comparison between the two codes for our nominal disc (Table \ref{tabledisc}). The blue dotted lines show the result using the planetesimal accretion code and the underlying red lines represent the results using the pebble accretion code. The outer most blue line at the top of the plot hides a red line below: they represent the initial profile. The disc evolves for 4.99 Myr and each line corresponds to the output each 100'000 years.}
\label{Tcomparison}
\end{figure}

\begin{figure}
\hspace{0cm} \includegraphics[height=0.35\textheight,angle=0,width=0.35\textheight]{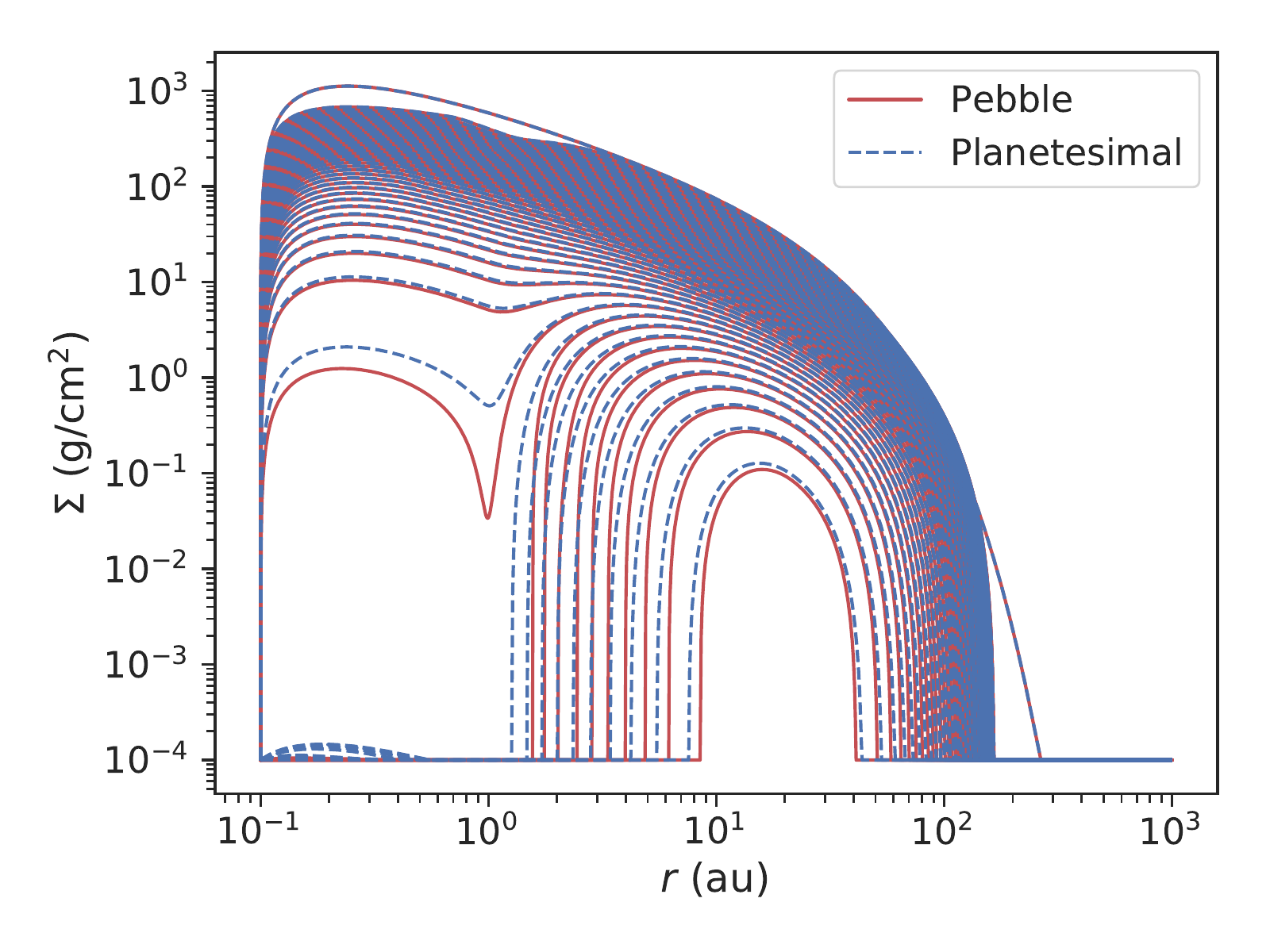}
  \caption{Surface density comparison between the two codes. Again here the red lines show the result using the pebble accretion code and the dotted blue lines give the outcome using the planetesimal accretion code.The disc evolves for 4.99 Myr and each line corresponds to the output each 100'000 years.}
  \label{sigmacomparison}
\end{figure}

Our first test case aims at comparing the evolution of the protoplanetary discs.
The same physical disc model \citep[following ][]{Hueso05} is used in both codes but since we use two distinct numerical implementations, a proper comparison is necessary to make sure that the same initial conditions lead to identical results.
Here, we focus on two quantities: the gas temperature and surface density. 
The temperature profile allows us to check that the vertical structure is giving identical results and the surface density is a key quantity for the formation of planets.
The simulations ran for 4.99 Myr, until the dissipation of the gas disc. The lines on Figs. \ref{Tcomparison} and Fig. \ref{sigmacomparison} represent the outcomes every $10^5$ years.
The outcome of the temperature comparison is represented in Fig. \ref{Tcomparison}, where we see the superposition of the temperature evolution in both codes. 
The results obtained using the pebble accretion code are hidden behind the results of the planetesimal accretion code.
They are indeed in very good agreement because they differ less than 1 \%. \\

The surface density comparison is shown in Fig. \ref{sigmacomparison}.
We see that the initial profile is exactly the same for both codes.
The physical description of the disc is identical in the two models.
However the numerics used to solve the equations are not implemented exactly the same way. 
Therefore, as the disc evolves, some divergences appear mainly after a few thousands of years of evolution.
The general agreement is however good: in the inner disc, the biggest difference we observe is 5 $\rm g/cm^2$, which is less than 1\%, and in the outer disc 1 $\rm g/cm^2$.
We can therefore conclude that the two discs evolve in a very similar fashion.

\subsection{Accretion of gas}
\label{accretionofgas}

We now consider a planet in the disc.
Its location is fixed at 5.2 AU to avoid the influence of migration.
We also set the solid accretion rate to $10^{-4}$ $\rm M_{\oplus}/ \rm yr$ to prevent the influence of how solids are accreted and to only compare the accretion of gas.
The initial mass of the core is $0.01$ $\rm M_{\oplus}$ and it is introduced in the disc after 0.1 Myr of evolution to allow the disc to reach a quasi-steady state.
To exclude any influence of the disc, we establish values for the planet boundary conditions that are fixed in time to make sure that the gas accretion and envelope structures are as similar as possible.
We choose a temperature $T$ of 60 K and a surface density $\Sigma$ at the planet location of 200 $\rm g/cm^2$, which are typical values for a location of 5.2 AU in a classic disc. \\

\begin{figure}
\hspace{0cm} \includegraphics[height=0.35\textheight,angle=0,width=0.35\textheight]{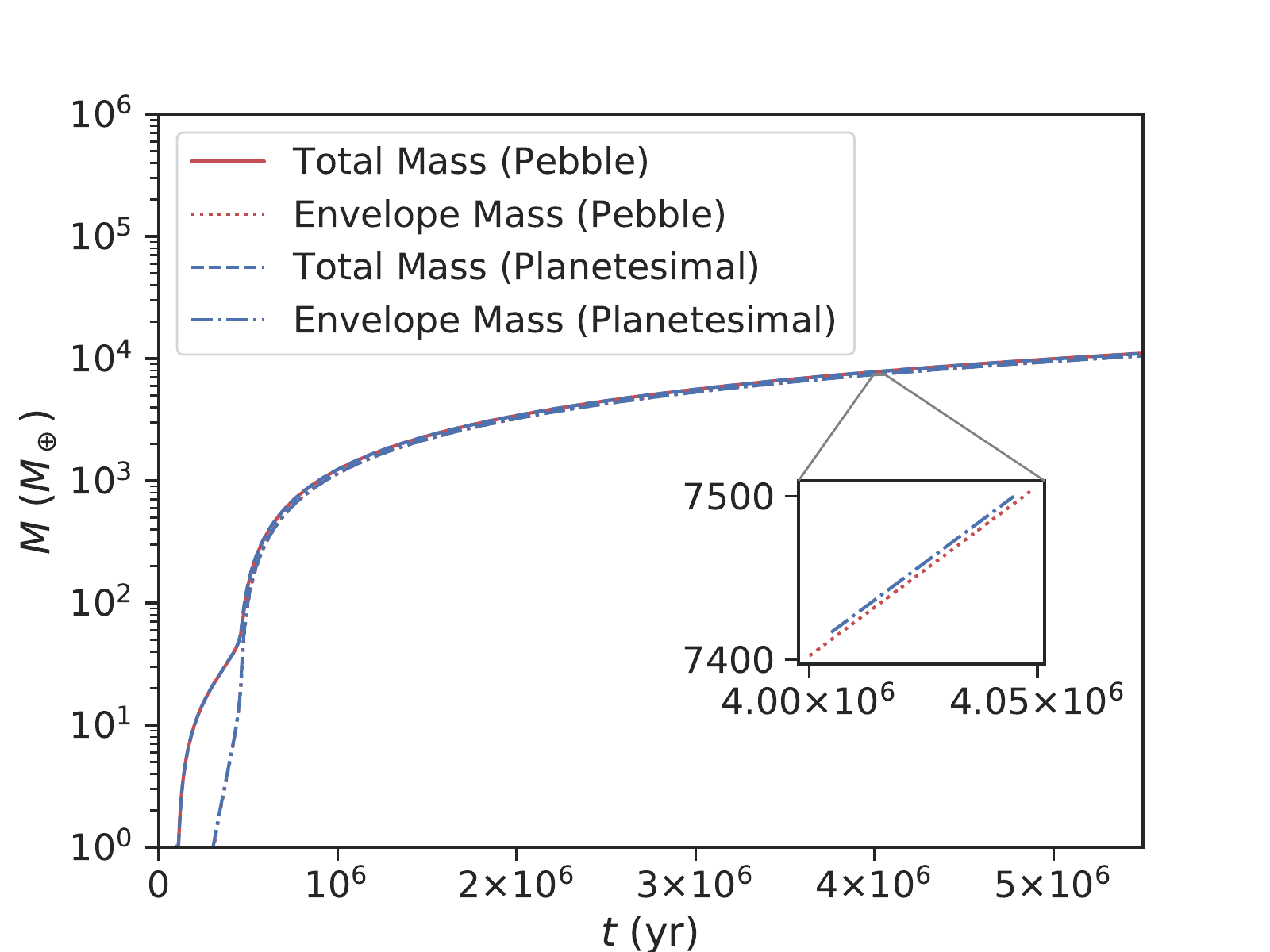}
  \caption{Gas accretion comparison for the two models. The outcomes of the pebble accretion model are shown in red: the solid line is the total mass and the dotted one is the envelope mass. The results of the planetesimal accretion code are represented in blue: the dashed line gives the total mass and the dashed-dotted line the envelope mass.The zoomed box helps understanding the behaviour of the envelope growth in a linear scale.}
  \label{Gasaccretioncomparison}
\end{figure}

As explained in Sect. \ref{gasaccretionmodel_theory}, the gas accretion rate onto the planet is given by the difference in envelope mass between two time-steps.
We however distinguish two regimes: when the planet is attached to the gas disc and when it undergoes runaway gas accretion.
In the second case, the accretion of gas is limited by what the disc can provide.
In Fig. \ref{Gasaccretioncomparison} we show a comparison of the gas accretion implementations.
The two envelope masses are represented as a function of time.
The previously mentioned runaway gas accretion phase starts, in our example, after $\sim 0.47$ $\rm Myr$ (see Fig. \ref{Gasaccretioncomparison}).
As shown in the zoomed area, the envelope masses are only differing by less than 0.1 \%.
We attribute this difference to the two distinct codes that may not converge to the exact same solution after the same number of iterations.

\subsection{Planet migration}
\label{planetmigraiton}

\begin{figure}
\hspace{0cm} \includegraphics[height=0.35\textheight,angle=0,width=0.35\textheight, trim={0 1.5cm 0 0}]{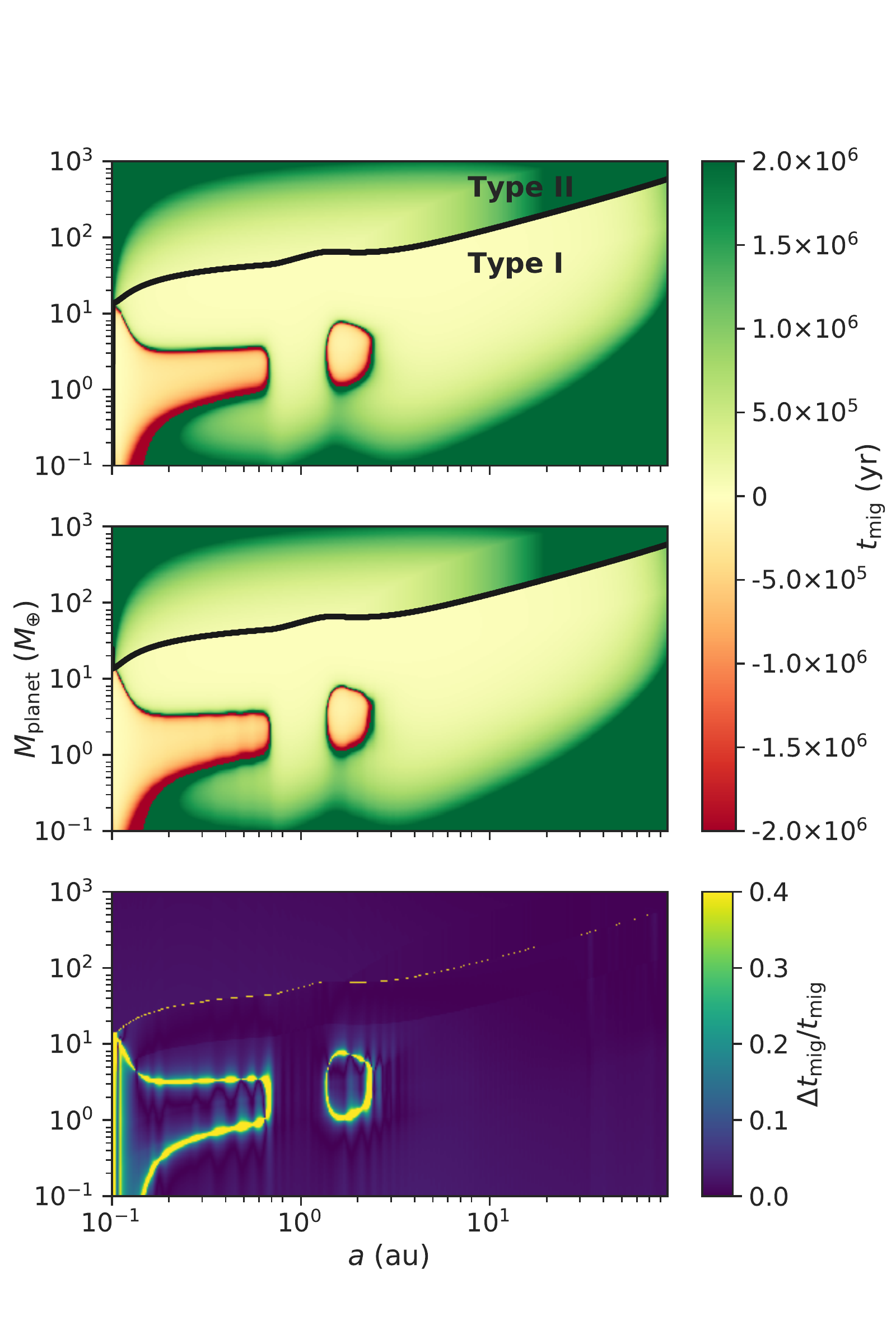}
  \caption{Map highlighting the different migration regimes after 0.1 Myr of disc evolution. In the upper two plots the red zones shows the outward migration regions. The type I and type II migration regime are distinguished by the solid black line: above it the planets migrate with type II migration and below they undergo type I migration. These two plots are computed with the planetesimal accretion code (most upper plot) and the pebble accretion code (middle plot) respectively. The third and bottom plot is the relative difference we observe between the two upper plots. The darker the outcome the more similar they are.}
  \label{Migrationmapcomparison}
\end{figure}

In our previous tests (see Sect. \ref{accretionofgas}), the planet location was fixed.
We now want to include the effect of migration because as they grow the planets migrate through the disc and the surrounding conditions are not identical at all locations.
It is therefore crucial to control that for a given scenario (fixed masses and identical initial locations), an embryo would follow the same path independent of the accretion model.
Using the same disc as previously introduced (see Table \ref{tabledisc}), our first comparison is in the form of a migration map to underline the migration regimes the planet may undergo. 
The maps are given in the upper two plots of Fig. \ref{Migrationmapcomparison} and are taken after 0.1 Myr of disc evolution.
The regions in red in these two plots indicate where the planet migrates outwards.
When located in the green areas the planet migrates inwards either through type I or type II migration depending on how massive they are.
The black line indicates the transition masses and locations between the two migration regimes.
In Fig. \ref{Migrationmapcomparison} the upper plot shows the migration map for the planetesimal accretion code, while the middle one shows the map for the pebble accretion code.
The bottom graph highlights the differences between the two outcomes: the darker the map, the more similar they are.
We observe two main differences: the first one along the outward migration regions and the second one along the inner edge of the disc. 
Even though it is not visible on the two upper plots, the outwards migration regions are shifted depending on the model.
These differences may be consequences of gradients that appear in the migration formulae for type I migration.
Indeed the surface density gradient, as well as the temperature gradient, are used in the computation of the Lindblad and corotation torques.
Computing gradients with two different solvers can thus  lead to divergences in the outcomes and the discs evolving slightly differently also impacts on the migration maps.  \\


\begin{figure}
\hspace{0cm} \includegraphics[height=0.35\textheight,angle=0,width=0.35\textheight]{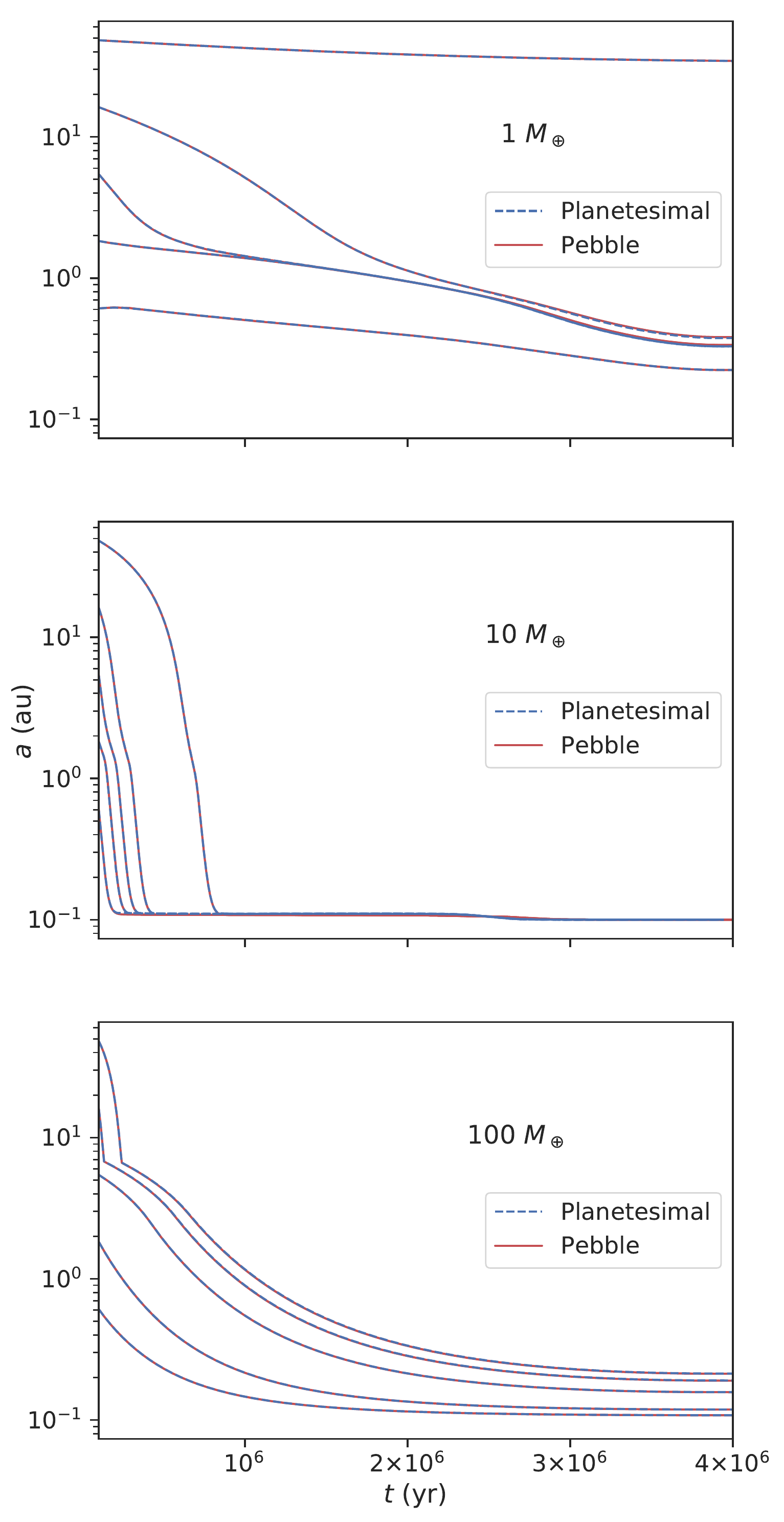}
  \caption{Migration of three different fixed-mass planets for different locations (0.6 AU, 1.8 AU, 5.5 AU, 17 AU and 50 AU). The upper plot shows the migration of a $1$ $\rm M_{\oplus}$ planet, the middle one a 10 $\rm M_{\oplus}$ planet and the bottom one a $100$ $\rm M_{\oplus}$ planet. The solid red lines give the outcomes of the pebble accretion code and the dashed blue lines represent the results of the planetesimal accretion code.}
  \label{Migration_fixedmassplanets}
\end{figure}

We then compare the migration of single planets with fixed masses.
In order to test different types of migration we use multiple initial locations and 3 distinct masses (1 $\rm M_{\oplus}$, 10 $\rm M_{\oplus}$ and 100 $\rm M_{\oplus}$) to  account for the three following migration regimes: type I, fast type I and type II respectively. 
The outcome of the comparison is shown in Fig. \ref{Migration_fixedmassplanets}.\\

In the upper plot we see the migration of a $1$ $\rm M_{\oplus}$ planet for different starting locations. 
As can be noticed in Fig. \ref{Migrationmapcomparison} (bottom plot), this particular mass lies in the region where the two outcomes of the codes differ the most, especially for locations below $1$ AU.
Furthermore the migration timescales (see the colour code in Fig. \ref{Migrationmapcomparison}, upper two plots) for a $1$ $\rm M_{\oplus}$ are the most diverse.
Indeed, depending on the location the planet may either migrate fast inwards or slowly inward, as well as outwards or experience zero migration regions.
Focusing first on the outermost planet, with an initial location of $\sim 50$ AU, it is in a region where the migration timescale is large, leading to a relatively slow migration. 
We therefore see that it remains near its initial location and end up around 40 AU after 4 Myr of disc evolution.\\

The planet starting at $\sim 18$ AU as well as the one starting at $\sim 6$ AU migrate relatively fast towards the inner edge of the disc until they reach $\sim 0.6$ AU where they cross a high migration time-scale region, leading to a slower migration regime.
This makes them stay nearly in the same location for 500'000 years.
The planet starting at $\sim 2$ AU experiences quite early on this slow migration regime as well and therefore ends up on a track similar to those of the two previous cases \citep{ColemanNelson16}.
When these three planets reach regions below 1 AU the two outcomes of the codes very slightly differ.
As we see in Fig. \ref{Migrationmapcomparison} (bottom plot), those are the regions where the outcomes of the codes differ the most, impacting here on the migration tracks.
As for the planet starting at $\sim 0.6$ AU, it starts further inside from the regions where the outcomes diverge and therefore the two tracks are matching each other.
This planet first experiences outward migration and then ends up in a zero migration area, which moves itself, making the final location of this planet only $\sim 0.3$ AU far from its original one.\\

In the center plot of Fig. \ref{Migration_fixedmassplanets}, the 10 $\rm M_{\oplus}$ planets experience fast type I migration.
Independent of their starting locations, they all migrate very quickly (less than 1 Myr) to the inner edge of the disc because their mass (10 $\rm M_{\oplus}$) lies in the range where type I migration is very efficient (see Fig. \ref{Migrationmapcomparison}, colour code of the upper two plots).
These planets are indeed not big enough to open a gap in the disc and therefore migrate with the type I regime, where the migration rate is proportional to the mass.
Furthermore the disc is dense at the begining of its evolution, which favours a rapid drift.
Comparing the behaviour of the planets for both models we get a very good agreement.\\

In the bottom plot of Fig. \ref{Migration_fixedmassplanets}, the migration of a 100 $\rm M_{\oplus}$ planet is presented.
Being more massive these planets usually open a gap and migrate in type II mode. 
Looking back at Fig. \ref{Migrationmapcomparison}, we see that a planet with a mass of 100 $\rm M_{\oplus}$ lies above the black line splitting type I and type II migration, meaning that it would migrate in type II for all locations below $\sim 20$ AU. 
The planets of the lower panel of Fig. \ref{Migration_fixedmassplanets} can then be split into two groups: the inner three planets and the two outer ones.
Looking at the three inner planets first, we see that they directly migrate in type II due to their mass and locations.
This prevents them from quickly migrating to the inner edge of the disc like the 10 $\rm M_{\oplus}$ planets.
It therefore takes them $\sim 2$ Myr to reach the inner edge even though they are initially located quite close to the star.
On the other hand the outer most planets first migrate in fast type I because of their location until they reach regions where they can undergo the type II regime leading to a slow migration towards the inner edge of the disc. 
Comparing the two models we again obtain very similar results.

\subsection{Combined effect of growth and migration}
\label{Combinedeffectofgrowthandmigraiton}

\begin{figure}
\hspace{0cm} \includegraphics[height=0.35\textheight,angle=0,width=0.35\textheight]{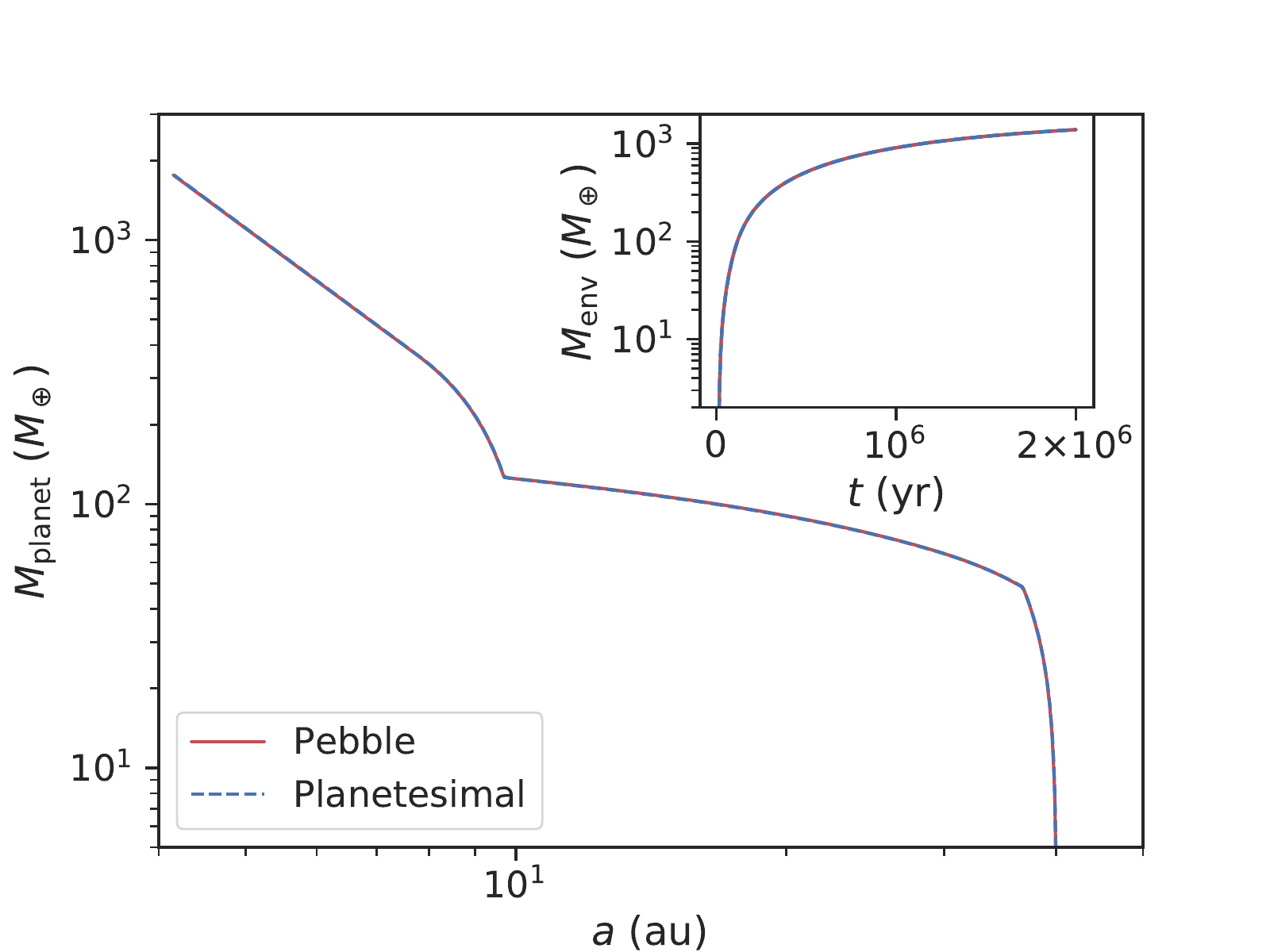}
  \caption{Mass of a migrating planet as a function of its location. A solid red line represents the pebble accretion code and is hidden behind the dashed blue line which gives the result of the planetesimal accretion code. The small window on the top right shows the mass of the envelope growing with time using the two codes. }
  \label{PlaneteMigcomparison}
\end{figure}

We now finally combine the effect of gas accretion and migration by looking at the mass growth of a single planet that migrates in a disc. 
For this test we use our nominal disc (Table \ref{tabledisc}), and insert a $0.01 M_{\oplus}$ planet at 40 AU at the beginning of the disc evolution.
As in the previous tests the accretion rate of solids is fixed to avoid any influence of the way solids are accreted (see Sect. \ref{accretionofgas}).
In order to trigger efficient gas accretion, we reduce exponentially the accretion rate of solids after 20 kyr of disc evolution.
The results are presented in Fig. \ref{PlaneteMigcomparison} where we see that the two codes give very similar results for the masses as a function of semi-major axis. 
The inset on the top right shows the temporal growth of the planet envelopes, which are also matching very well. 
This test is the closest to a real simulation we could produce without any impact of the solid accretion models.
Given the excellent similarity between the results in this test we can now explore the effects of the two solid accretion models knowing that the other components of the computation are very similar and will not induce differences.

\section{Population synthesis outcomes}
\label{populationsynthesisoutcomes}

\subsection{Initial conditions}
\label{Initialconditions}

We use the nominal model outlined in Sect. \ref{planetmigraiton}, with the disc model being similar to that of \citet{Hueso05} and described in Sect. \ref{discmodel_theory}.
The accretion of gas onto the planet follows the equations introduced in Sect. \ref{gasaccretionmodel_theory}.
The opacity of the planetary envelope is reduced by a factor $f_{\rm opa} = 0.003$ because observations hint that the grain opacity is smaller than the full interstellar one \citep{Mordasini14}.
The accretion of solids differs between the two models: either planetesimals with radii of \unit[1]{km} (see Sect. \ref{planetesimalaccretionmodel}) or pebbles (mm to cm size) are accreted (see Sect. \ref{pebbleaccretionmodel}).
While growing, as explained in Sect. \ref{planetmigration_theory}, the planet interacts with the disc and starts migrating through the disc.

We run simulations of single planet per disc to avoid the chaotic effects of N-body simulations and allow a proper comparison of the two models.
The embryo is inserted at different times of the disc evolution ($0$ Myr, $0.2$ Myr, $0.5$ Myr and $1$ Myr) to explore the impact on the resulting populations (10'000 planets per starting time).
Its location is randomly chosen from a uniform distribution in logarithmic space between 0.1 and 50 AU and the initial mass of this inserted body is 0.01 $\rm M_{\oplus}$.

With our populations we aim at taking a wide range of discs into account. 
We randomly draw masses from the distribution of inferred Class I gas disc masses by \citet{Tychoniec18} and multiply them with another random value drawn from the distribution of spectroscopic metallicities obtained by \citet{Santos05} to obtain dust disc masses. 
The exponential cut-off radius of the gas disc profile is a function of the gas disc mass following \cite{Andrews10} and the cut-off radius for the planetesimal disc (where applicable) is half of the former \citep{Ansdell18}. 
The subsequent disc evolution is then governed by $\alpha$ (see Table \ref{tabledisc}) and photo-evaporation (see Sect. \ref{discmodel_theory}). 
To have disc lifetimes matching the lifetime distribution inferred from observations of disc fractions in stellar clusters \citep[e.g.][]{Mamajek09}, we linearly scale the external photo-evaporation by a third random number drawn from a log-normal distribution \citep[see][]{Mordasini15}.
The total amount of solids is randomly drawn from the aforementioned distribution, but from this amount of solids, part of it forms the bodies that can be accreted, while the rest remains as dust. We present here two scenarios of how the total solid mass is distributed: either 90\% forms the accretable bodies with 10\% of the mass in dust ($\varepsilon = 0.9$ case) or 50\% forms the accretable solids with 50\% remaining dust ($\varepsilon = 0.5$ case).

For purpose of simplicity for the comparison we do not use here the full versions of the two models (as would be done for example for the planetesimal accretion model in Emsenhuber et al. in prep).
For instance for both the pebble and the planetesimal accretion models the radius of the solid core of the planet is calculated using a fixed density of \unit{5.5}{ g/cm$^3$}, which is a simplification compared to what is used by \citet{Mord12a,Mord12b}.
This facilitates the analysis by avoiding second order effects on the gas accretion via an otherwise emerging core contraction luminosity.
The potential feedback of the composition of the accreted solids is therefore lost. 
For this reason, we only track the composition in terms of silicates and water ice.
The separation of the icy and rocky population, given by the water ice line, is calculated using the midplane pressure and temperature at the starting time of the simulation.\\

We stress that for a detailed comparison with observations, the interactions between the growing planets are important \citep{Alibert13} and the populations presented here are intended to simulate realistic conditions for the different solid accretion mechanisms, but are not meant to be compared to the observed population of planets. We leave this for future studies.\\

\subsection{Mass vs semi-major axis}
\label{Massvssemimajoraxis}

\begin{figure*}[!ht]
\makebox[\linewidth]{
\begin{tikzpicture}
		\node (img) {\includegraphics[clip,width=\linewidth]{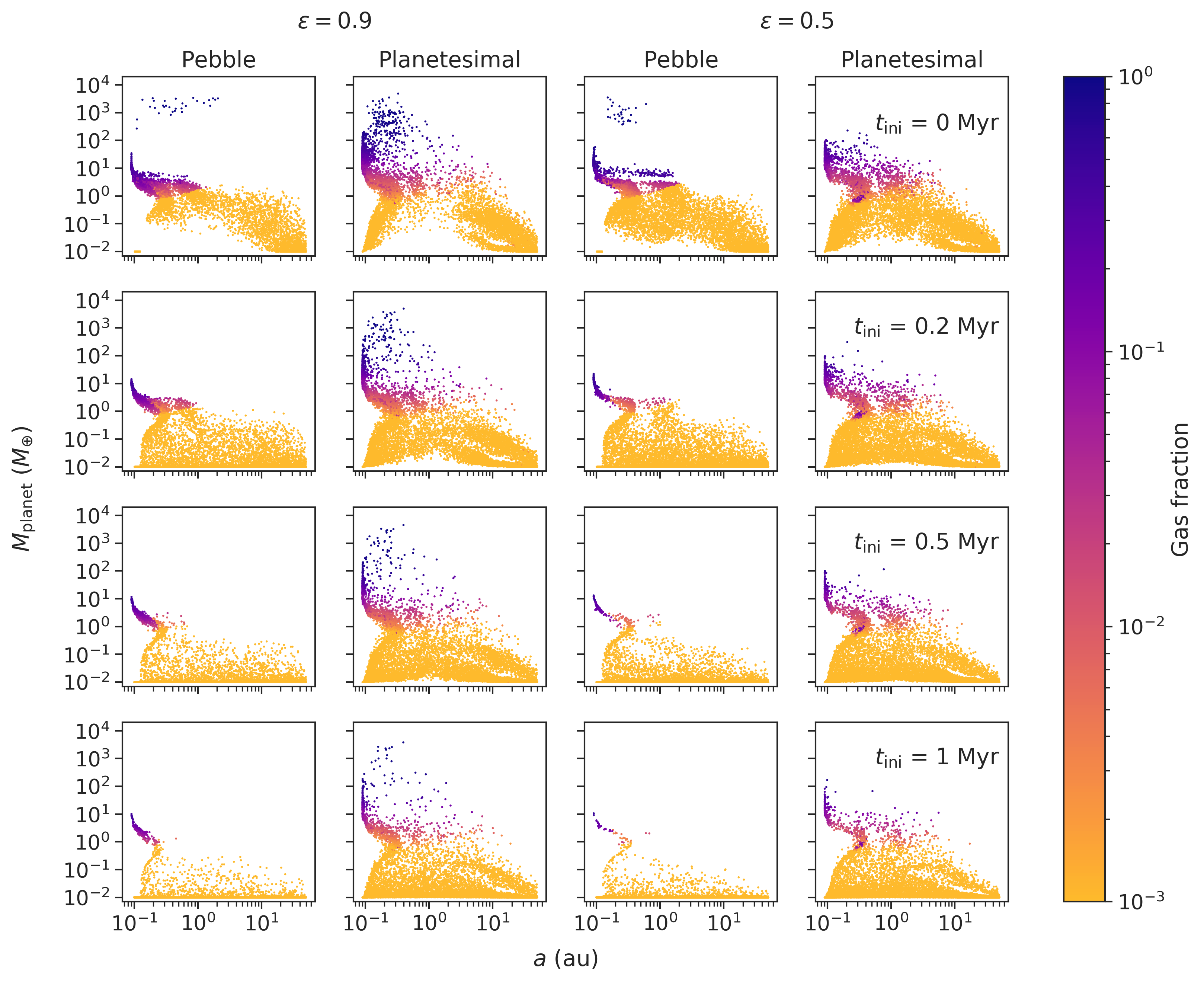}};
\end{tikzpicture}}
  \caption{Mass as a function of semi-major axis of all planets in all populations using $\varepsilon = 0.9$ (left two columns) or $ \varepsilon = 0.5$ (right two columns). In these two blocks the left column always gives the results for the pebble accretion model and the right one the product of the planetesimal accretion model. Each line represents a starting time. The colour code expresses the gas fraction for each planet at the end of the formation stage.}
  \label{MvsA_9050percent_gasfraction}
\end{figure*}

Fig. \ref{MvsA_9050percent_gasfraction} shows the mass of the formed planets as a function of their final locations for different starting times. 
The two columns on the left differ from those on the right by the amount of solids used to form the bodies that can be accreted by the planet ($\varepsilon$).
Within these two partitions the respective left column of the panels gives the output for the pebble accretion model and on its right are the results for the planetesimal model.
The colour code expresses the gas fraction of each planet.
Focusing first on the case where $\varepsilon = 0.9$ (left two columns of Fig. \ref{MvsA_9050percent_gasfraction}) we immediately see that different types of planets are formed by the two models.
Using the planetesimal accretion model, more giant planets\footnote{We consider that a giant planet is a planet with a mass higher than 100 $\rm M_{\oplus}$.} are produced than with the pebble model, independent of the starting time.
The pebble model indeed only produces giants for the $\rm t_{\rm ini} = 0$ Myr case.
For this specific starting time it also only produces very few planets with masses between $\sim 80$ $\rm M_{\oplus}$ and $\sim 1000$ $\rm M_{\oplus}$ compared to the planetesimal scenario.
Finally most of the giants, albeit very few in number, are very massive.
This is due to the fact that planets growing by pebble accretion only start accreting gas efficiently when solid accretion is stopped. 
Thus if the planets have a massive enough core and are located in the outer disc, they may undergo type II migration and have time to accrete a considerable gaseous envelope.

A general behaviour observed for both models is that the starting time impacts the mass of the formed planets: the earlier the embryo is inserted the more massive the planets.
The variability in the starting times however impacts the planets formed by pebble accretion more.
Indeed the growth of the planets in the pebble model depends on the pebble front.
This growth front is the place where the dust particles have grown to pebble size and start migrating towards the star.
It moves outwards with time and induces a pebble flux.
When the pebble front reaches the outermost radius of the disc, the pebble flux drops to zero.
If this happens at times earlier than $t_\mathrm{ini}$ then no growth occurs.
In the model the time at which the growth radius reaches the outer edge of the disc scales with the metallicity and can therefore strongly vary. 
The average time is however around $\sim 300'000$ years.
Therefore, especially for later starting times, some planets do not grow at all because there is no flux of pebbles anymore (see the bottom panels of Fig. \ref{MvsA_9050percent_gasfraction}). 
This starting time effect has less impact in the planetesimal model, where some growth is always possible, unless the planet is located very far away from the star where the planetesimal accretion rates are extremely small, or there are no planetesimals in the planet's feeding zone. 

Another important feature in the $t_\text{ini} = 0$ case is the faster growth inside the snowline in the pebble model compared to the planetesimal one.
The Stokes number of pebbles is reduced when crossing the ice line because of ice sublimation \citep{IdaGuillot16}.
This impacts on the accretion rate of pebbles, which is divided by a factor $\sim$ 2 \citep{Lodders2003}.
However, even with this accretion reduction, the pebble flux reaching an embryo located inside the ice line remains significantly larger compared to the planetesimal accretion rate on an embryo inside the ice line in the same disc. 
 The planetesimal rate is considerably reduced in these regions due to the proximity to the star and the resulting smaller feeding zone, which is a function of the Hill radius.
The semi-major axis versus mass distribution of the intermediate mass planets is in all cases distinctly shaped by migration, as can be seen by the over-densities of planets in regions of outward migration that are clearly visible.\\

Moving to the two right columns of Fig. \ref{MvsA_9050percent_gasfraction}, where $\varepsilon = 0.5$, some general conclusions drawn for the $\varepsilon = 0.9$ case also apply: the transition in the envelope masses occurs for smaller masses using the planetesimal model and the early starting times help to form more massive planets.
The amount of giant planets formed by the planetesimal accretion model is however strongly reduced compared to the $\varepsilon = 0.9$ case.
This is caused by the decrease in the available solids to form the massive cores that are needed to grow into giants. 
For the pebble model going from $\varepsilon = 0.9$ to $\varepsilon = 0.5$ does not have such a dramatic impact on the abundance of giant planets.
The abundance of giants is indeed more impacted by the pebble isolation mass, which acts as a threshold for the planet to reach larger masses.
If the planets do not reach $\rm M_{\rm iso}$, they won't accrete an envelope, independently of the $\varepsilon$ value. 
However the general tendency for both models in the $\varepsilon = 0.5$ case compared to the $\varepsilon = 0.9$ case is that the planets are less massive (this will be further discussed in Sect. \ref{Distribution}).\\

Taking a closer look at the colour code we see that the transition between practically no envelope (orange dots) and a small envelope (pink dots) looks different in the two models.
While for the planetesimal model the transition between a total solid core (orange dots) and a body with a small envelope (pink dots) is smooth, in the pebble formation model, we see a clearer distinction.
This is due to the gas accretion starting only when the planets reach the pebble isolation mass.
The distinction we see, which has a diagonal shape between $\sim0.2$ and 2 AU for masses between $\sim 1$ and 5 $\rm M_{\oplus}$, is therefore an imprint of $\rm M_{\rm iso}$.
Focusing on the $\varepsilon = 0.5$ case for the planetesimal model, we see a few planets with masses around $1$ $\rm M_{\oplus}$ and semi-major axis between 0.2 and 0.4 AU that have higher gas mass fractions (see the concentration of purple dots while the background is orange in the right column of Fig. \ref{MvsA_9050percent_gasfraction}).
These planets experience outward migration and, since they already emptied their feeding zone, start accreting gas as soon as their semi-major axis increases.
In the $\varepsilon = 0.9$ case we do not see this feature appearing because the planets were massive enough to accrete a more significant envelope.\\



\begin{figure}
\hspace{0cm} \includegraphics[height=0.35\textheight,angle=0,width=0.35\textheight]{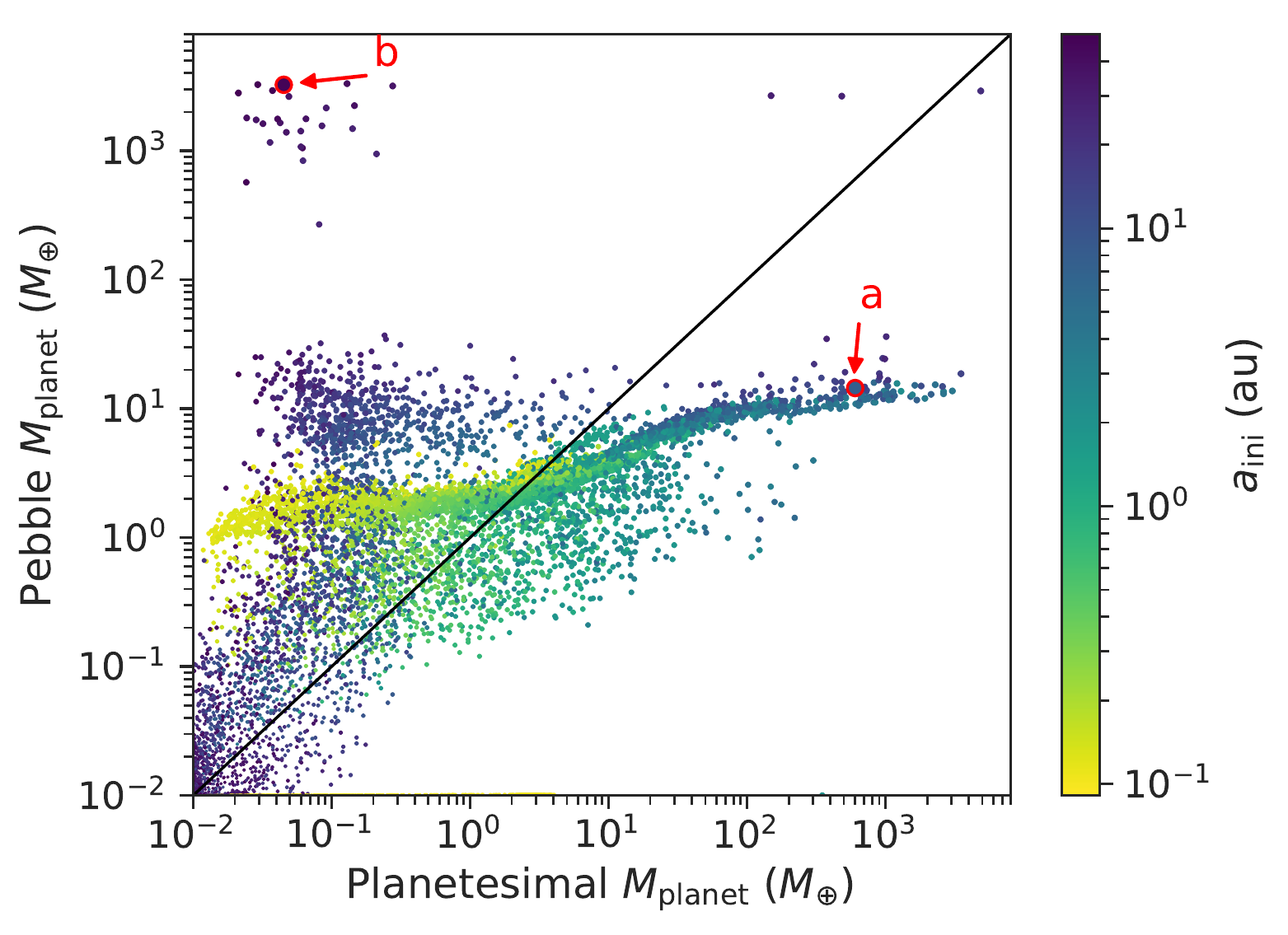}
  \caption{Mass of the planets formed by pebble accretion as a function of the mass formed by planetesimal accretion for our nominal case ($\varepsilon = 0.9$, $\rm t_{\rm ini} = 0$ Myr. The colour code gives the initial location of the embryos. The two red circle indicate the two cases that are discussed in Sect. \ref{Growthtracks}. The point size is scaled with the $\rm max(M_{p, \rm peb}, M_{p, \rm plan})$ for better visibility in the small mass ranges. }
  \label{MpMp}
\end{figure}


In Fig. \ref{MpMp} we focus on the situation where the embryos are inserted at $\rm t_{\rm ini}=0$ Myr for the $\varepsilon=0.9$ case and define it as our nominal case.
The outcomes in terms of mass of the population of planets formed by pebble accretion are represented as a function of the population formed by planetesimal accretion.
The colour code gives the initial location of the embryos.
We clearly see on this plot that the giant planets formed by planetesimal accretion remain around super-Earth masses in the pebble accretion model.
We also see through the colour code that these planets initially formed between $\sim 1$ to 10 AU.
On the other hand planets starting further out did not grow much in the planetesimal model, while they reach 10 to 30 $\rm M_{\oplus}$ when growing by pebble accretion. 
Some of them even grow into giant planets (see the planets on the top left of the figure).
This hints on the impact that the starting locations have in both models: growing in the inner disc is more favourable to planetesimal accretion while starting in the outer regions of the disc is beneficial to pebble accretion.
Indeed in the outer regions of the disc $\rm M_{\rm iso}$ is larger, allowing the planets growing by pebble accretion to have more massive cores, which can trigger efficient gas accretion.
This may lead to gap opening and prevent the planet from being lost to the star.

\subsection{Populations analysis}
\label{Distribution}

\begin{figure}
\hspace{0cm} \includegraphics[height=0.35\textheight,angle=0,width=0.35\textheight]{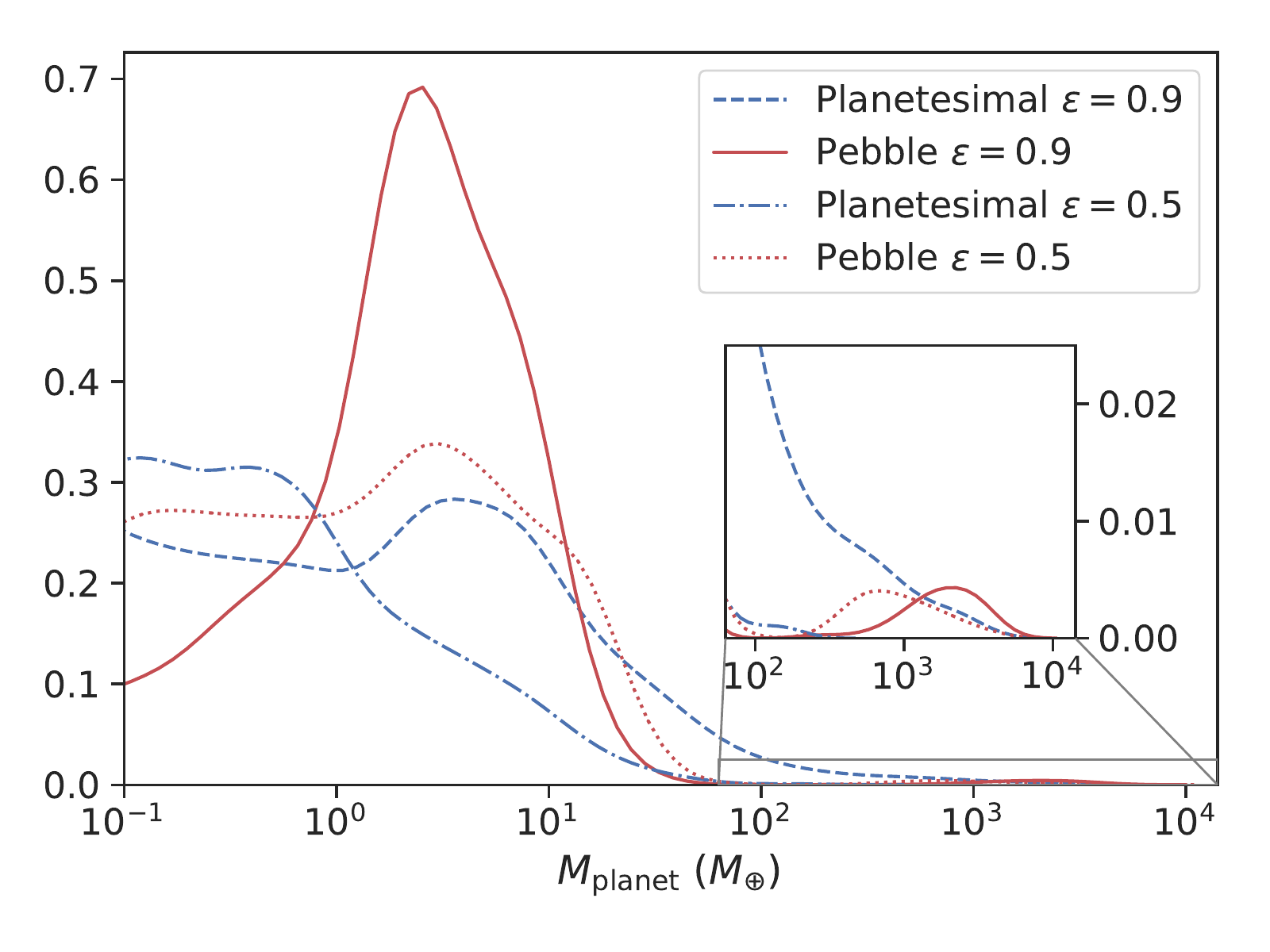}
  \caption{Kernel density estimate for starting time $\rm t_{\rm ini}= 0$ Myr. The red lines are for the pebble model and the blue lines for the planetesimal model. We show here the results for the two partitions of the amount of solids:  solid and dashed lines are used when $\varepsilon = 0.9$ while dotted and dashed-dotted lines are used when $\varepsilon = 0.5$ .The kernel density estimates were obtained using a Gaussian kernel with a Normal reference rule bandwidth \citep{Scott92}.}
  \label{KDEplot}
\end{figure}

In order to further compare the two accretion models and especially increase the visibility in the overpopulated regions of the scatter plots, we present the same results with mass distributions.
We focus on the case where $\rm t_{\rm ini}=0$ Myr (Fig. \ref{MvsA_9050percent_gasfraction}, top line) because it is the case where the pebble model is able to form giant planets.
In Fig. \ref{KDEplot} we look at the types of planets formed depending on the partition of solids: either $\varepsilon =0.9$ or $\varepsilon = 0.5$. 
Looking first at the red lines (pebble model) we see that the $\varepsilon = 0.9$ case (solid line) forms more super-Earth mass planets while the $\varepsilon = 0.5$ (dotted line) case forms less massive planets.
This is due to the lack of solid material available for accretion by the embryos.
However more 50 $\rm M_{\oplus}$ planets form in the $\varepsilon = 0.5$ case because the planets grow more slowly and therefore, if they reach $\rm M_{\rm iso}$, they do it at a later stage of the disc evolution, when migration is less efficient.
This gives them more time to accrete gas while migrating towards the star.
In the $\varepsilon = 0.9$ case they did not accrete as much gas and migrated into the star.
Comparing the amount of giant planets (in the zoomed box) we see that there is a shift in the masses but the total number of these types of planets is still relatively low.
The decrease in the amount of pebbles therefore mainly acts on the less massive planets for the pebble model.

For the planetesimal accretion model (blue lines) we obtain similar results to the pebble model. 
With $\varepsilon = 0.5$ (dashed-dotted line), there are more low mass planets and less super-Earth mass planets. 
The amount of giant planets, however, strongly decreases compared to the $\varepsilon = 0.9$ (dashed line) case because there are fewer available solids to form planetesimals and a large amount of planetesimals is needed to form giants.\\

Fig. \ref{KDEplot} also provides information to compare the two models with each other. 
We focus on the $\varepsilon = 0.9$ case.
First, we clearly see that the behaviours of the two lines are slightly shifted but both show a bump around super-Earth mass planets.
The pebble model however forms more of them compared to the planetesimal model.
This is due to the isolation mass: for low $\rm M_{\rm iso}$, when the solid accretion is stopped, gas accretion remains very slow. 
Therefore these super-Earth mass planets do not accrete large envelopes and stay in the mass range where type I migration is efficient.
They thus migrate into the inner 1 AU of the disc and then get trapped at zero migration regions, migrating with them as the regions migrate over time \citep{ColemanNelson14}.
This results in planets not accreting a significant gaseous envelope and consequently remain at super-Earth masses. 
In the planetesimal model on the other hand the planets continue to accrete solids while they start accreting gas.
The transition between solid and gas accretion is therefore more smooth.
This helps growing to larger masses than super-Earth because the accretion onto the planets depends on the Hill radius, and thus the more massive the planets, the larger their Hill radii, the more they accrete.
Additionally the onset of gas accretion increases the planetesimal capture radius \citep{Inaba} which leads to further growth.

We highlight the larger masses in a zoomed area on the right of the plot, which helps in comparing the amount of giant planets. 
As mentioned in Sect. \ref{Massvssemimajoraxis}, the pebble model does not produce many planets between $\sim 80$ $\rm M_{\oplus}$ and $\sim 1000$ $\rm M_{\oplus}$ compared to the planetesimal model.
Going back to Fig. \ref{MpMp} this hole in the mass range of the planets formed by pebble accretion is even clearer, while in the planetesimal accretion case we see that all types of masses form.
This is due to the fact that if a planet becomes massive enough, it crosses the fast type I migration bottleneck by opening a gap in the disc and can then migrate with type II migration, which is much slower than type I. 
The rare giant planets in the pebble case are bound to become very massive because they reach this regime earlier when there is still a lot of gas to accrete. 
Planets growing by planetesimal accretion reach type II migration over a larger range of times.
Therefore it results in a larger spread in final masses for the giant planet population.
When looking at larger masses, both models predict the formation of some very massive planets ($> 1000$ $ \rm M_{\oplus}$).
Additionally, the decrease in the numbers of super-Earths to Neptunes is much sharper in the pebble model because of the very few planets with masses between 80 and 1000 $\rm M_{\oplus}$.

\begin{figure}
\hspace{0cm} \includegraphics[height=0.35\textheight,angle=0,width=0.35\textheight]{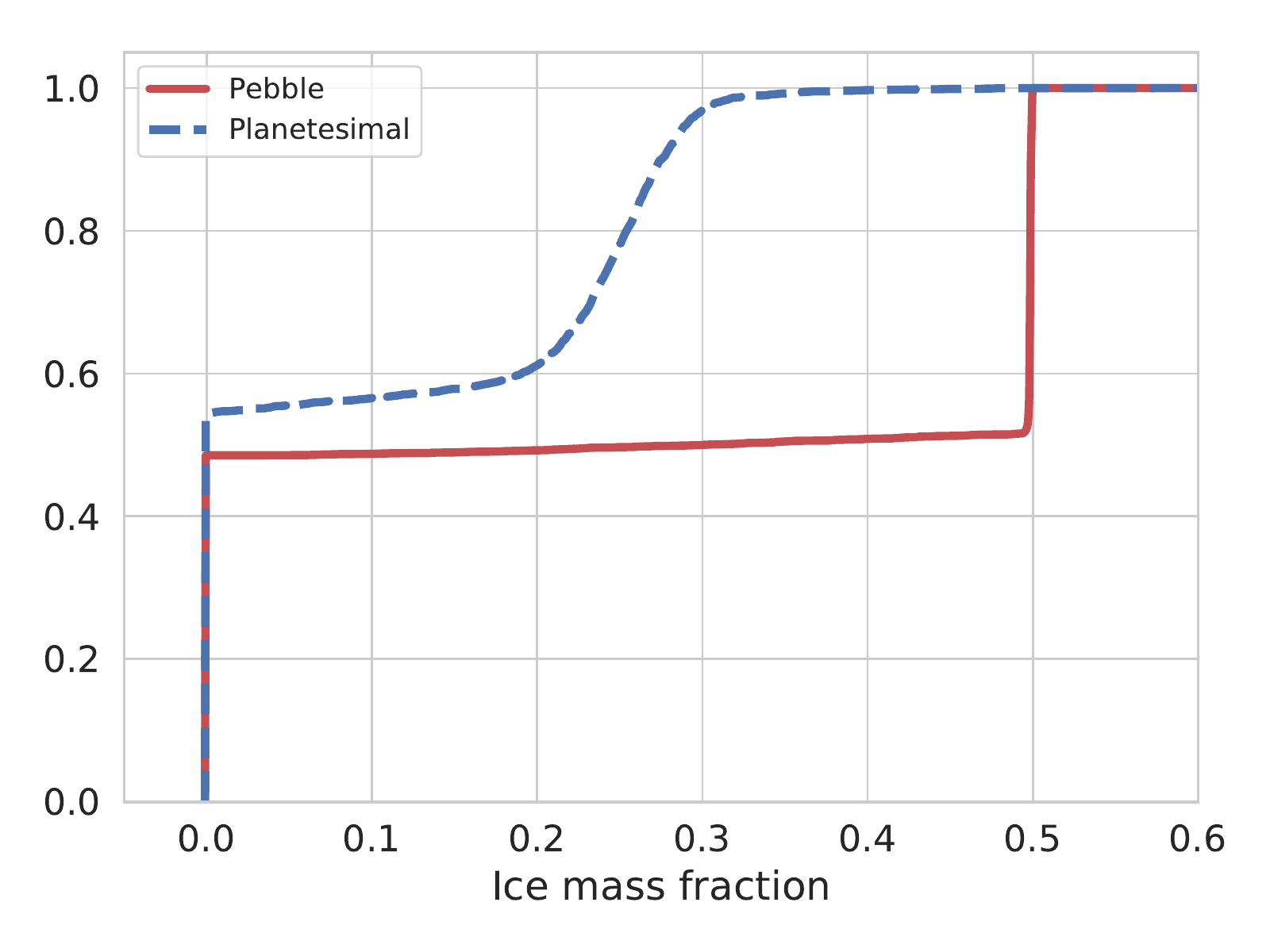}
  \caption{Cumulative distribution of the ice mass fractions in the solid core of the planets for our nominal case ($\varepsilon = 0.9$, $\rm t_{\rm ini} = 0$ Myr). The red line provides the pebble accretion model results while the blue one gives the planetesimal accretion model results. We focus here on masses higher than 1 $\rm M_{\oplus}$ and semi-major axis below 1 AU.}
  \label{Icemassfraction}
\end{figure}

We now look at some final properties of the formed bodies.
In Fig. \ref{Icemassfraction} we provide a cumulative distribution of the ice mass fractions for our nominal case ($\varepsilon = 0.9$ and $\rm t_{\rm ini} = 0$ Myr).
We focus on bodies with masses higher than 1 $\rm M_{\oplus}$ and orbits inside 1 AU to take into account planets that may be observed by transit measurements.
We therefore concentrate here on bodies that are mainly composed of solids and do not discuss the amount of water in the envelopes.
We consider the bodies (planetesimals or pebbles) accreted by the embryo outside the ice line to be composed of 50\% ice and 50\% rock and the embryo itself as well, if formed outside the ice line \citep{Lodders2003}.
If these bodies are accreted inside the ice line, the ice sublimates and therefore the solids are only made of rock.
The same applies for the embryo, if initially located inside the ice line, it is 100\% rocky.
The two models produce quite different results.
Focusing on the red line first (pebble model) we see that either the embryo is fully rocky, or made of 50\% ice and 50\% rock. 
There are barely any planets with an intermediate composition. 
This is due to the fast accretion of solids: pebbles are very efficiently accreted by the growing embryo and therefore the accretion of solids mainly occurs near the initial location, before any migration of the forming planets.
The location where the planet reach $\rm M_{\rm iso}$ is indeed on average more than 80\% alike the initial location of the planet.
Furthermore the migration of the ice line is negligible over the time the embryo accretes pebbles.
This "in-situ" solid accretion results in solid cores that are either completely formed outside the ice line or completely formed inside.
Barely any embryo migrates during its solid accretion phase to be able to obtain an intermediate composition.  
Computing the same figure for $\varepsilon = 0.5$ or for a later $\rm t_{\rm ini}$ would not impact on the sharp profile of the ice compositions. 
However it would increase the amount of planets with a solid composition only. But the sharp transition between a solid composition and a 50\% ice composition would remain because of the fast growth by pebble accretion compared to the ice line migration timescale.

For the planetesimal model (blue line) we also focus on bodies with masses higher than 1 $\rm M_{\oplus}$ and orbits inside 1 AU and find that the rocky bodies are dominant.
Compared to the pebble model their abundance is even higher.
No planets have a 50\% ice and 50\% rock composition unlike the pebble model because the forming planets have a slower growth and start migrating while accreting solids.
This impacts on the intermediate compositions.
$\sim 40$ \% of the planets have ice mass fractions between $\sim 0.05$ and $\sim 0.25$.
Because the planetesimal accretion rate is lower than the pebble one, the growth of the core takes more time.
Therefore the growing embryos start to migrate while accreting planetesimals, allowing them to cross the ice line while accreting solids, resulting in reduced ice fractions.
\citet{Schoonenberg19}, as well as \citet{Coleman19}, discuss the theoretical water content of the planets in the frame of the Trappist-1 system.
Combining the effect of planetesimal and pebble accretion \citet{Schoonenberg19} obtain a water fraction of the order of 10 \%.
This result is closer to the planetesimal accretion scenario result we obtain in the present work.
As for the values we present for planets formed by pebble accretion, they are in agreement with \citet{Coleman19}.

\begin{figure}
\hspace{0cm} \includegraphics[height=0.35\textheight,angle=0,width=0.35\textheight]{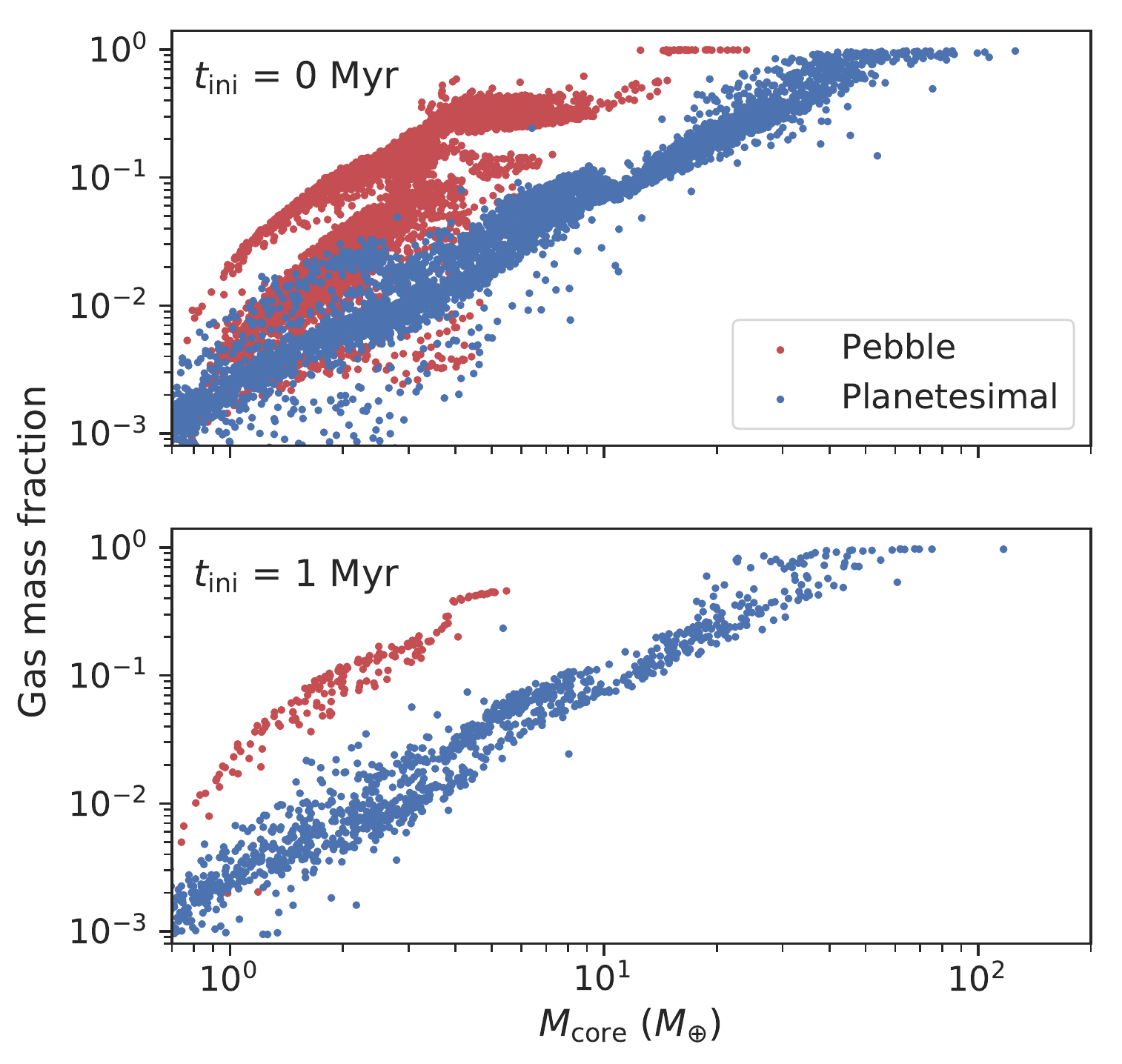}
  \caption{Gas mass fraction as a function of the core mass for the $\varepsilon = 0.9$ case. Again here the pebble model results are in red, while the planetesimal model outcomes are in blue. The upper plots shows the results for our nominal case ($\rm t_{\rm ini}=0$ Myr) and the bottom one for $\rm t_{\rm ini}=1$ Myr. We therefore see here that the starting time of the embryo does not impact on the general outcome.}
  \label{Gasmassfraction}
\end{figure}

Another interesting result is the distribution of gas mass fractions. 
We represent this distribution as a function of the core mass in Fig. \ref{Gasmassfraction} for our nominal case as well as for $\rm t_{\rm ini} = 1$ Myr.
We see that the envelope fraction for a given core is generally higher using the pebble accretion model.
The two plots underline that the results are alike and therefore independent of $\rm t_{\rm ini}$.
The divergence is due to the components of the luminosity of the planets that strongly differ in the two models.
At this stage of the formation the total luminosity is dominated by its solid accretion component because the cores mainly accrete solids.
Thus when planets formed by pebble accretion reach their isolation mass and stop accreting solids, the solid accretion luminosity is strongly reduced.
This therefore induces an increase of the gas accretion luminosity which will engender efficient gas accretion (in agreement with \citealp{Alibert2018}).
On the other hand, at the same formation stage, planets growing by planetesimal accretion continues to accrete planetesimals whilst simultaneously accreting gas, which supplies considerable solid accretion luminosity.
This leads to a smaller gas accretion luminosity and therefore less gas accretion.
This translates here in higher gas mass fractions for a given core mass using the pebble accretion model.\\

\begin{figure}
\hspace{0cm} \includegraphics[height=0.35\textheight,angle=0,width=0.35\textheight]{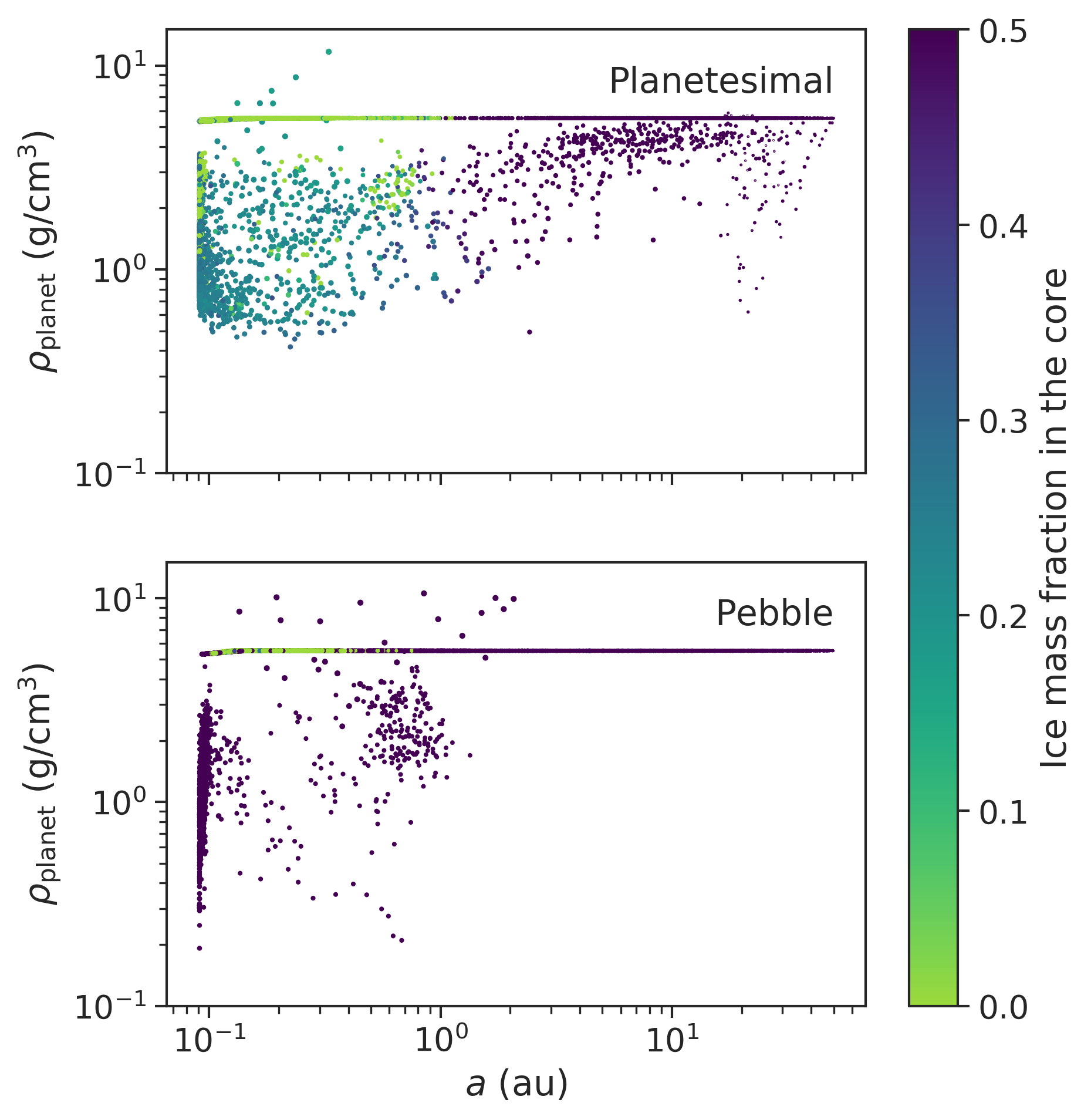}
  \caption{Density of the planet as function of the final location of the planets for the $\varepsilon = 0.9$ case and starting time $\rm t_{\rm ini}=0$ Myr. The colour code expresses the ice mass fraction in the core. The upper plots provides the results of the planetesimal accretion model and the bottom one the ones of the pebble accretion model.}
  \label{Densityplot}
\end{figure}

As gas mass fractions are not directly observable we take a look at the resulting densities and whether the differences between the models are still present after the long-term evolution phase (Sect. \ref{Longtermevolution})
The composition of the planets is a good indicator of the differences between the two models.
Fig. \ref{Densityplot} highlights these divergences with the density of planets represented as a function of their final location.
The colour code gives the ice mass fraction in the core to indicate the core composition.
The horizontal line we see on both plots for a density of $5.5$ $\rm g/ \rm cm^3$ is an imprint of the fixed core density we use in our models.
Therefore these planets have practically no envelopes and were represented with orange dots in Fig. \ref{MvsA_9050percent_gasfraction}.
In Fig. \ref{Densityplot} the colour code is impacted by the ice line: if the planets grow inside the ice line, they are mainly rocky and are therefore represented by green dots, while if they grow further outside they have a 50 \% ice composition and are characterised by purple dots.
The colour code description is indeed similar to that provided for Fig. \ref{Icemassfraction}: there is no planet with intermediate compositions for the pebble model, while the planetesimal model shows many of them.\\

Fig. \ref{Densityplot} shows that for locations beyond $\sim 3$ AU, the pebble model only predicts rocky bodies with density of $5.5$ $\rm g/ \rm cm^3$.
This means that these planets do not have an envelope.
This is an imprint of our pebble accretion model where gas can only be accreted once the planets reach the isolation mass.
This is very different from the planetesimal accretion model where we see many planets located outside 3 AU with density smaller than $5.5$ $\rm g/ \rm cm^3$ which means that they accreted an envelope.
This feature was also visible in the two top left panels of Fig. \ref{MvsA_9050percent_gasfraction}, which represent our nominal cases. \\

What is however interesting to point out with Fig. \ref{Densityplot} is the outcomes for planets located inside 3 AU.
There both models predicts the formation of planets with envelopes. 
Therefore most of them have densities smaller than $5.5$ $\rm g/ \rm cm^3$.
In the pebble accretion model the high gas mass fractions we obtain in Fig. \ref{Gasmassfraction} even lead to some very low density planets ($\rho < 0.5$ $\rm g/cm^3$).
The planetesimal accretion model is not forming planets with such low densities.
This means that if these intermediate mass gas-rich planets would be observed, the pebble accretion scenario could help understanding their formation.\\
On the other hand some very massive planets (> 1000 $\rm M_\oplus$) have densities larger than $5.5$ $\rm g/ \rm cm^3$.
This is due to the decrease in radius that happens with such high masses \citep{Mord12b}.
Focusing on these dense planets we discuss first the ones formed by planetesimal accretion.
They have intermediate core compositions because they accreted solids while migrating and crossed the ice line as they grew.
Furthermore they reach such high masses because when they accrete gas it augments the colisional probability \citep{Inaba} and therefore also increases the solid accretion (see further discussion in Sect. \ref{Growthtracks}).
On the other hand the dense planets formed by pebble accretion all have a 50\% ice composition because they accreted all their solid material outside the ice line.
Their growth in the outer disc was quick and nearly in-situ, and since the isolation mass is bigger in these regions of the disc, they formed massive cores.
These massive cores lead to efficient gas accretion and helps in forming very massive planets, leading to these high densities.


\subsection{Growth tracks}
\label{Growthtracks}

\begin{figure}
\hspace{0cm} \includegraphics[height=0.35\textheight,angle=0,width=0.35\textheight]{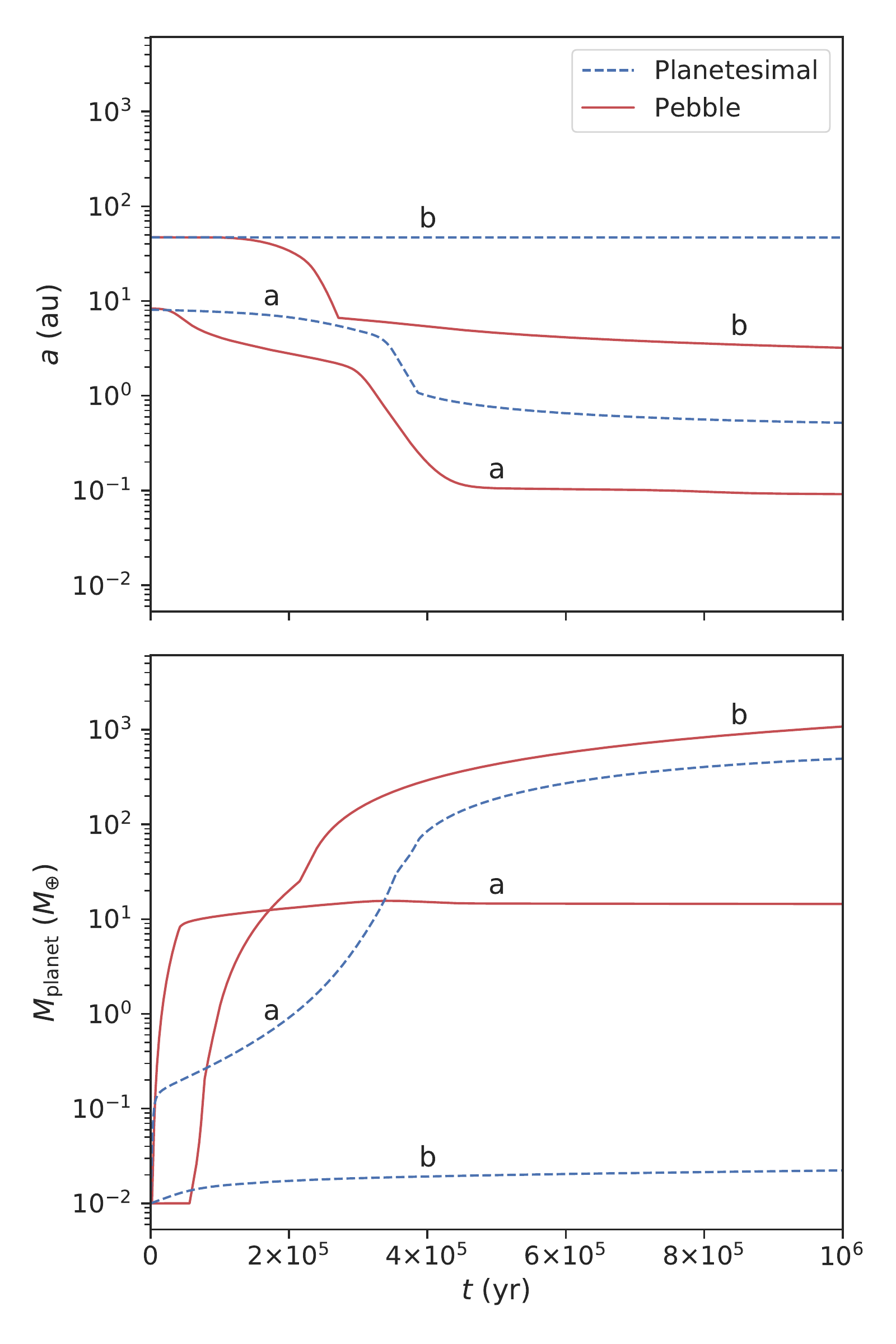}
  \caption{Growth tracks and migration tracks as a function of time for two disc cases. In the disc $a$ the planetesimal accretion model forms a giant planet while in the disc $b$ the pebble accretion model does. The upper plot shows the migration of the planets with time and the bottom plot the mass of the planets with time. Again the red lines represent the pebble model results and the blue lines the planetesimal model results.}
  \label{tracksInTime_comparison}
\end{figure}

One of the conclusions of the previous section is that the type of planets formed differs between the two models.
For instance we saw in Fig. \ref{KDEplot} that hardly any giants formed through the accretion of pebbles. 
To illustrate the different formation path we look at two different disc cases (see Fig. \ref{tracksInTime_comparison}).
In the first disc case, disc $a$, the planetesimal accretion model forms a giant planet.
We compare its tracks with the the ones of the planet formed by pebble accretion that grew in the same disc.
The initial conditions for the two simulations of disc $a$ are therefore the same.
In the disc $b$, the pebble model forms a giant planet.
We also compare its tracks with the ones of the planet formed by planetesimal accretion for the same initial conditions.
The outcomes of the two cases are highlighted with red circles in Fig. \ref{MpMp}.
The disc initial conditions are similar between disc $a$ and $b$, except for the photoevaporation rate, which impacts on the disc lifetime.
Disc $b$ has a slightly longer lifetime because the photoevaporation rate is smaller than the one in disc $a$.
The top panel of Fig. \ref{tracksInTime_comparison} shows the migration of the planets with time while the bottom one provides the masses as a function of time.\\

We focus first on disc $a$ where the planetesimal model forms a giant planet.
The initial location of the planet is $\sim 8$ AU.
For the pebble case (disc $a$, red line) we see that since the planet grows very rapidly, it starts migrating efficiently early in its evolution.
The planet indeed quickly reaches $\sim 10$ $\rm M_{\oplus}$ by only accreting solids while type I migration has a big impact on its location.
Thus, when it is massive enough to accrete gas efficiently, it is already around 2 AU and therefore continues to migrate to the inner edge of the disc without any time to accrete a considerable envelope.
Approaching the inner edge of the disc then hampers the accretion of gas on to the planet since the planet's Hill sphere is significantly reduced, and small amounts of gas accretion are sufficient to supply the luminosity generated by the planet \citep{CPN17}.
On the other hand, the embryo formed through planetesimal accretion (disc $a$, blue line) grows more slowly because of the lower planetesimal accretion rate compared to the pebble one.
Furthermore it is only once the planet starts accreting gas efficiently (just before $4 \times 10^5$ years) that  inward migration becomes important.
Because the planet is already quite massive ($\sim 40$ $\rm M_{\oplus}$) it accretes gas efficiently and quickly opens a gap in the disc.
This allows the planet to start migrating in the type II regime, which is slower than type I, resulting in the formation of a giant planet.

Looking now at disc $b$, where the pebble model forms a giant, we see that the outcomes are very different for the two models.
The initial location for these planets is $\sim 47$ AU.
Focusing first on the planet formed by pebble accretion (disc $b$, red line) we see that it does not start accreting pebbles at the very beginning of the simulation. 
The planet mass remains constant for $\sim 50'000$ years.
This is a consequence of the growth front of pebbles ($r_g$) not reaching the location of the embryo before this time.
When $\rm r_g$ finally reaches the initial location of the planet, the latter grows and rapidly attains its isolation mass ($\sim 20$ $\rm M_{\oplus}$).
In the meantime the planet migrates with type I migration, but since the growth to the isolation mass is quite rapid, the fast migration does not occur for too long to be an issue for the future planet growth.
Thus the planet is massive enough to accrete gas efficiently while undergoing type II migration. 
The planet therefore accretes its large envelope while slowly migrating towards the central star and ends up forming a giant planet located around 2 AU.
In these regions of the disc the accretion of planetesimals is more difficult than the one of pebbles.
Indeed the pebble surface density in these locations is higher than the planetesimal one because the planetesimal surface density profile is steeper than the one of the gas disc (see Sect. \ref{planetesimalaccretionmodel}, \citealp{Drazkowska17,Lenz19}) while the pebble surface density profile undergoes a similar slope to the one of the gas.
Additionally, planetesimal accretion becomes very inefficient due to the collisional timescale increasing with the semi-major axis.
The result thus shows that the planet formed by planetesimal accretion (disc $b$, blue line) does not grow much and remains as a failed core \citep{Mords09} near its initial location. 
The two results are divergent and we see that the starting locations plays an important role in the outcomes of the simulations within the two accretion models, as we already discussed with Fig. \ref{MpMp} in Sect. \ref{Massvssemimajoraxis}.\\

\begin{figure}
\hspace{0cm} \includegraphics[height=0.35\textheight,angle=0,width=0.35\textheight]{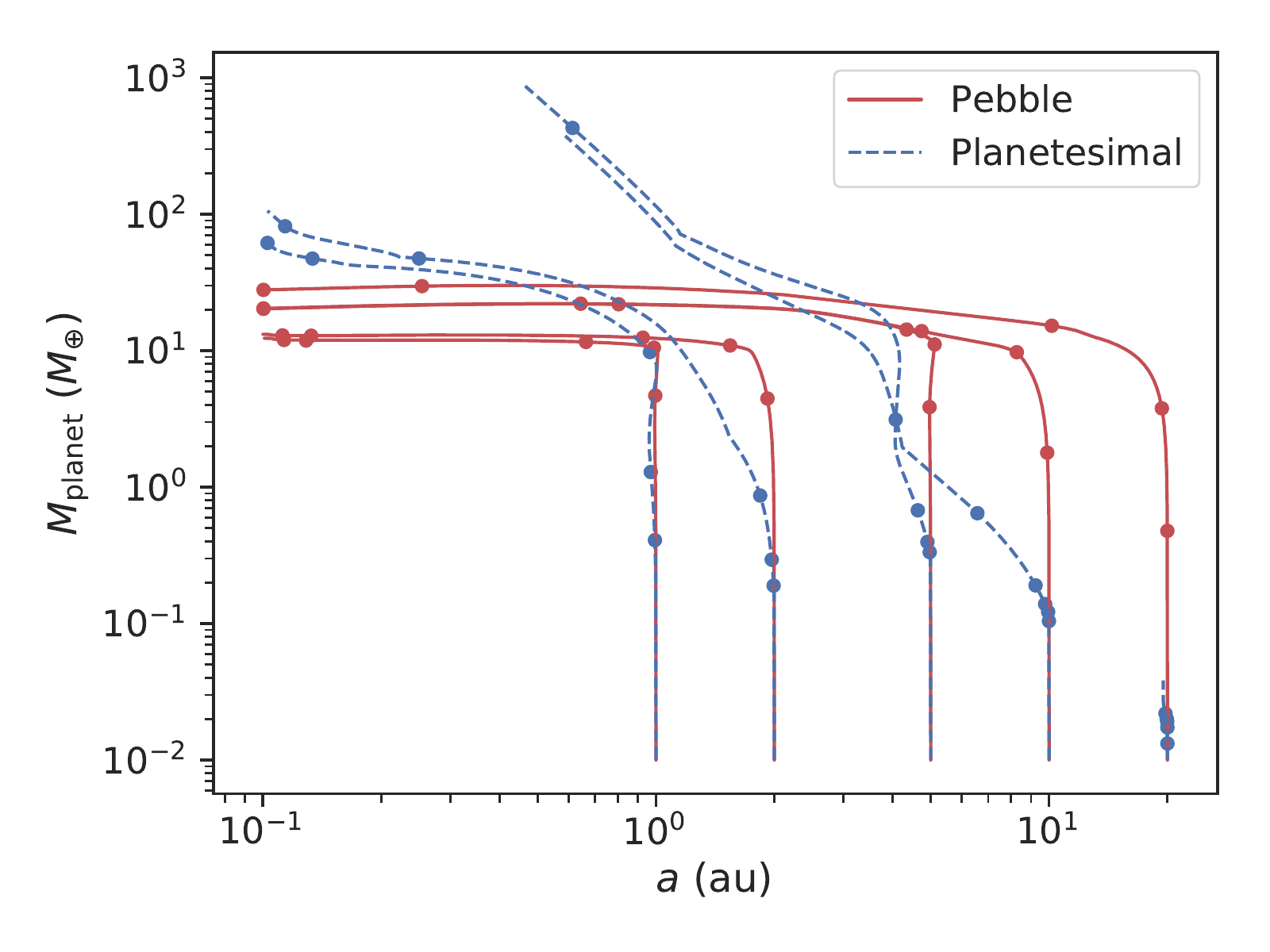}
  \caption{Growth tracks of planets in the same disc for a starting time of $\rm t_{\rm ini}= 0$ Myr. The initial locations are 1, 2, 5, 10 and 20 AU. The red solid lines give the pebble model outputs while the blue dashed lines represent the planetesimal results. The dots indicates the growth evolution after 10'000 years, 30'000 years, 0.1 Myr, 0.3 Myr and 1 Myr. The disc lifetime for this simulation is 2 Myr.}
  \label{Growthtracks_0Myr}
\end{figure}

In order to gain a feeling of how planets behave depending on their initial locations we use our nominal disc (given in Table \ref{tabledisc}) and increase the initial amount of solids to $\rm Z = 0.1$ to ensure growth and choose different starting locations for the planets (1, 2, 5, 10 and 20 AU).
Starting the embryos at 0 Myr we obtain the growth tracks provided in Fig. \ref{Growthtracks_0Myr}.
The red lines give the pebble accretion model results while the blue ones represent the planetesimal accretion model outcomes.
We immediately see that the two models produce very different tracks -- as already concluded with Fig. \ref{tracksInTime_comparison}.
However, we see that using the pebble model, type I migration is very efficient and planets with masses around $\sim10$ $\rm M_{\oplus}$ migrate directly to the inner edge of the disc without any chance of accreting significant amounts of gas.
These planets indeed reach their isolation masses rapidly (see the red dots on the tracks, indicating the time evolution), essentially growing almost in-situ.
They are then not massive enough to trigger efficient gas accretion which can aid them in avoiding fast type I migration.
On the other hand, what prevents planets formed by planetesimal accretion to also have this behaviour is that they accrete solids more slowly (see the blue dots on the tracks, indicating the time evolution) and start migrating while accreting solids.
Consequently they reach the outward migration regions ($\sim 1$ AU and $\sim 3$ AU), which prevents them from directly falling into the star.
 This puts them in a favourable location given their mass to accrete gas efficiently. 
 Additionally, when they accrete gas, the planetesimal accretion rate increases due to gas drag that enlarges the collisional probability described by \cite{Inaba}. 
 This leads to a significant increase in solid accretion while gas accretion is also occuring.  
This is in strong contrast to the pebble accretion model where, when planets reach the pebble isolation mass and gas accretion becomes efficient, the solid mass does not increase anymore.


 \section{Discussion and conclusion}
 \label{conclusion}

This work provides a comparison between two planet formation scenarios: pebble accretion and planetesimal accretion. 
Using two distinct codes we utilise the same disc model, gas accretion model and migration model.
A proper testing can only be done if the initial conditions are identical, which is why we compare the implemented disc model, the accretion of gas and the migration regimes.
The comparison yielded convincing results (Fig. \ref{PlaneteMigcomparison}), allowing the two solid accretion models to be adequately compared.\\

Using a population synthesis approach we then compute simulations of single planet per disc to avoid the chaotic effects of the use of an N-body integrator. 
We leave the interactions between several planets in a common disc for future work.
The embryos in our simulations are inserted at different starting times (0, 0.2, 0.5 and 1 Myr) with initial locations uniform in logarithm between 0.1 and 50 AU.
We choose two scenarios to split the amount of solids available in the disc: either we use 90\% of this amount to form the accretable bodies (planetesimals or pebbles), or we use 50\%.
The rest of the solids remains as dust and contributes to the opacity of the disc.
For the envelope calculation the grain opacity is reduced by a factor $\rm f_{\rm opa} = 0.003$ because grain-free opacities are more relevant for the envelope calculation than interstellar medium-like opacities \citep{Mordasini14}. This reduction is applicable to our pebble accretion model because the planets formed by pebble accretion only accrete solids for masses below $\rm M_{\rm iso}$. At this point, they have practically no envelope because of the high solid accretion luminosity that prevents gas accretion. This therefore avoid for the pebbles to evaporate inside the envelope and for them to impact on the grain opacity.
We investigated the influence of a change of opacity in the envelope for planets below $\rm M_{\rm iso}$ and did not obtained a significant impact.\footnote{Note that in our previous work \citep{Brugger18}, we reduced the opacity also above $\rm M_{\rm iso}$, which is why there was a change in the mass functions.}

A general observation is that the outcome of the populations (Fig. \ref{MvsA_9050percent_gasfraction}) is very different depending on the accretion model. 
Indeed the planetesimal accretion model forms a larger amount of giant planets.
The pebble model produces a few giants mainly when the embryo is inserted at $\rm t_{\rm ini}=0$ Myr and they are very massive (more than a thousand Earth masses).
The starting time indeed has a big impact for the pebble accretion model. 
The earlier the embryo is inserted, the more massive the planets.
Furthermore for later starting times, some planets may not grow at all because of the absence of pebble flux when the pebble front reaches the outer edge of the disc.
On the other hand, for the planetesimal accretion scenario, the initial starting time plays a less important role since growth is possible at any time.
The growth of the planets in the planetesimal model is more influenced by the location because if the planet is located far away from the star, planetesimal accretion rates are extremely low. 

The impact of the splitting in the amount of solids appears mainly when using the planetesimal accretion model (Fig. \ref{KDEplot}).
The amount of giant planets is clearly reduced when less solids are available to form the planetesimals since a large amount of planetesimals is needed to grow large cores.
Figure \ref{KDEplot} reveals a gap in the mass distribution of planets around Jupiter masses when using the pebble accretion model.
Comparing the mass distributions of the two models, more super-Earth mass planets are formed by the pebble model and the decrease from super-Earth to Neptunes is much sharper in this model as well because of the very few planets with masses between 80 and 1000 $\rm M_{\oplus}$ formed by pebble accretion.\\

We then compare the ice mass fractions (Fig. \ref{Icemassfraction}) and see that using the pebble accretion model the resulting planets are either fully rocky or with a 50\% rock 50\% ice composition. 
Few intermediate compositions are formed because the planets grow fast and nearly in-situ.
On the other hand the planetesimal model does produce a significant number of planets with intermediate compositions because the planets formed by planetesimal accretion grow slower while migrating.

Focusing on the gas mass fraction (Fig. \ref{Gasmassfraction}) we find that the pebble model forms planets with higher gas mass fractions for a given core mass compared to the planetesimal model. 
We show that this result is independent of the starting time of the embryo but is influenced by the contributions to the luminosity of the planets.
Planets formed by pebble accretion have a low solid accretion luminosity once they reach their isolation mass because solid accretion is stopped.
This results in a high gas accretion luminosity \citep{Alibert2018}, which triggers efficient gas accretion \citep{CPN17}.
Planets formed by planetesimal accretion on the other hand accrete gas while still accreting solids, leading to a lower gas accretion luminosity.
This finally translates into gas mass fractions for a given core mass that are higher for planets formed by pebble accretion.
This difference in the gas mass fractions is not always retained over the Gyr evolution after the disc dispersal due to efficient photo-evaporation of the atmosphere. 
However some differences between the models are still present after the long-term evolution. Intermediate mass planets formed by pebble accretion indeed reach densities as low as 0.2 $\rm g/cm^3$, where the lower limit reached by planets formed by planetesimal accretion is 0.5 $\rm g/cm^3$ (Fig. \ref{Densityplot}).
Therefore only the pebble model could form such gas-rich intermediate mass planets.\\

However the planets formed by pebble accretion do not grow to giants because of their too efficient migration (see Fig. \ref{tracksInTime_comparison}).
They indeed grow quickly to their isolation mass and therefore reach in the early evolution of the disc the mass range where type I migration is decisive.
Migration is very efficient in a dense disc and the planet reaches the inner regions of the disc very quickly, without enough time to accrete a significant envelope \citep{ColemanNelson14,ColemanNelson16}.
On the other hand, planets formed by planetesimal accretion have a slower growth rate.
Furthermore when they start accreting gas, they are still accreting solids, which slows their accretion of gas due to an increased solid accretion luminosity \citep{Alibert2018}.
Thus the transition between pure solid accretion and gas accretion is not abrupt, helping them to become massive enough to open a gap in the disc.
This therefore reduces their migration rate, allowing them more frequently to accrete gas efficiently and grow to giant planets.\\

The lack of giant planets formed by the pebble accretion model is interesting to compare with \citet{Brugger18}.
A substantial amount of giants was obtained when reducing the opacity in the planetary envelope. 
However in \citet{Brugger18} the disc profile is different (as mentioned in Sect. \ref{discmodel_theory}), leading to a higher surface density in the outer regions and therefore a higher flux of pebbles. 
The variability in the amount of solids (ranging between 0.011 and 0.11) is also a key factor for the formation of giants.
In the present work we focus on a distribution with a mean value around $\rm Z_{\rm tot} = 0.02$, which lies in the lower range of what was used in the previous work and therefore the amount of giant planets is affected.
For similar amounts of solids, the same types of planets as in the present work were obtained.

For the planetesimal accretion model the amount of giant planets is impacted by the size of the planetesimals.
We focused in this work on 1 km sized planetesimal, which helps the formation of giants compared to bigger sizes \citep{Fortier13}.
This highlights that both scenarios require specific conditions to form giants.
A hybrid approach \citep{Alibert2018} might help to overcome the difficulties linked to each model.

Our work underlines the impact of the different accretion scenarios: pebble accretion or planetesimal accretion.
We should however keep in mind that we focus on single planet populations and therefore the consequences of mutual interactions between the planets are not taken into account.
We leave this improvement for future studies.

\acknowledgements

This work has been carried out within the frame of the National Centre for Competence in Research PlanetS supported by the Swiss National Science Foundation. The authors acknowledge the financial support of the SNSF under grant 200020\_172746.

\bibliography{references}
\bibliographystyle{aa}

\end{document}

%% file: def.tex
\def\f1{f_{\rm I}}

\def\beq{\begin{equation}}
\def\eeq{\end{equation}}

\def\simgr{\,\hbox{\hbox{$ > $}\kern -0.8em \lower 1.0ex\hbox{$\sim$}}\,}
\def\simle{\,\hbox{\hbox{$ < $}\kern -0.8em \lower 1.0ex\hbox{$\sim$}}\,}
\def\beq{\begin{equation}}
\def\eeq{\end{equation}}

\def\simgr{\,\hbox{\hbox{$ > $}\kern -0.8em \lower 1.0ex\hbox{$\sim$}}\,}
\def\simle{\,\hbox{\hbox{$ < $}\kern -0.8em \lower 1.0ex\hbox{$\sim$}}\,}
\def\beq{\begin{equation}}
\def\eeq{\end{equation}}

\def\apjl{ApJ}                
\def\apjs{ApJS}               
\def\ao{Appl.Optics}          
\def\aap{A\&A}                

\def\({\left(}
\def\){\right)}
\def\<{\left<}
\def\>{\right>}

\def\({\left(} 
\def\){\right)} 
\def\<{\left<} 
\def\>{\right>} 

\def\bc{\begin{changebar}}
\def\bce{\begin{center}}
\def\beq{\begin{equation}} 

\def\bi{\begin{itemize}}

\def\btab{\begin{tabular}{p{1.7cm}p{12cm}p{1.5cm}}}
\def\bt2{\begin{tabular}{p{1 cm}p{4.5cm}p{10cm}}}

\def\ec{\end{changebar}}
\def\ece{\end{center}}
\def\eeq{\end{equation}} 
\def\ei{\end{itemize}}

\def\etab{\end{tabular}\\}

\def\mH2{m_\mathrm{H_2}}

\def\dmax0{\rho_\mathrm{max}}
\def\dmaxS0{\Sigma_\mathrm{max}}

\def\rH2{r_\mathrm{H_2}}

\def\r0max{r_\mathrm{0max}}

\def\s0{\sigma_\mathrm{0}}

\def\xp0{x_{\rm{M0}}}

\def\z0max{z_\mathrm{0max}}

\def\apjl{ApJ}                
\def\apjs{ApJS}               
\def\ao{Appl.Optics}          
\def\aap{{\it A\&A}}                

%% file: Planetesimal_pebble.bbl
\begin{thebibliography}{67}
\expandafter\ifx\csname natexlab\endcsname\relax\def\natexlab#1{#1}\fi

\bibitem[{{Adachi} {et~al.}(1976){Adachi}, {Hayashi}, \& {Nakazawa}}]{Adachi}
{Adachi}, I., {Hayashi}, C., \& {Nakazawa}, K. 1976, Progress of Theoretical
  Physics, 56, 1756

\bibitem[{{Adams} {et~al.}(1988){Adams}, {Lada}, \& {Shu}}]{Adams88}
{Adams}, F.~C., {Lada}, C.~J., \& {Shu}, F.~H. 1988, \apj, 326, 865

\bibitem[{{Alexander} \& {Pascucci}(2012)}]{AlexanderPascucci12}
{Alexander}, R.~D. \& {Pascucci}, I. 2012, \mnras, 422, 82

\bibitem[{{Alibert}(2016)}]{Alibert16}
{Alibert}, Y. 2016, \aap, 591, A79

\bibitem[{{Alibert} {et~al.}(2013){Alibert}, {Carron}, {Fortier}, {Pfyffer},
  {Benz}, {Mordasini}, \& {Swoboda}}]{Alibert13}
{Alibert}, Y., {Carron}, F., {Fortier}, A., {et~al.} 2013, \aap, 558, A109

\bibitem[{{Alibert} {et~al.}(2005){Alibert}, {Mordasini}, {Benz}, \&
  {Winisdoerffer}}]{Alibert05}
{Alibert}, Y., {Mordasini}, C., {Benz}, W., \& {Winisdoerffer}, C. 2005, \aap,
  434, 343

\bibitem[{{Alibert} {et~al.}(2018){Alibert}, {Venturini}, {Helled}, {Ataiee},
  {Burn}, {Senecal}, {Benz}, {Mayer}, {Mordasini}, {Quanz}, \&
  {Sch{\"o}nb{\"a}chler}}]{Alibert2018}
{Alibert}, Y., {Venturini}, J., {Helled}, R., {et~al.} 2018, Nature Astronomy,
  2, 873

\bibitem[{{Andrews} {et~al.}(2010){Andrews}, {Wilner}, {Hughes}, {Qi}, \&
  {Dullemond}}]{Andrews10}
{Andrews}, S.~M., {Wilner}, D.~J., {Hughes}, A.~M., {Qi}, C., \& {Dullemond},
  C.~P. 2010, \apj, 723, 1241

\bibitem[{{Ansdell} {et~al.}(2018){Ansdell}, {Williams}, {Trapman}, {van
  Terwisga}, {Facchini}, {Manara}, {van der Marel}, {Miotello}, {Tazzari},
  {Hogerheijde}, {Guidi}, {Testi}, \& {van Dishoeck}}]{Ansdell18}
{Ansdell}, M., {Williams}, J.~P., {Trapman}, L., {et~al.} 2018, \apj, 859, 21

\bibitem[{{Ataiee} {et~al.}(2018){Ataiee}, {Baruteau}, {Alibert}, \&
  {Benz}}]{Ataiee2018}
{Ataiee}, S., {Baruteau}, C., {Alibert}, Y., \& {Benz}, W. 2018, \aap, 615,
  A110

\bibitem[{{Bell} \& {Lin}(1994)}]{Bell94}
{Bell}, K.~R. \& {Lin}, D.~N.~C. 1994, \apj, 427, 987

\bibitem[{{Birnstiel} {et~al.}(2012){Birnstiel}, {Klahr}, \&
  {Ercolano}}]{Birnstiel12}
{Birnstiel}, T., {Klahr}, H., \& {Ercolano}, B. 2012, \aap, 539, A148

\bibitem[{{Bitsch} {et~al.}(2015){Bitsch}, {Johansen}, {Lambrechts}, \&
  {Morbidelli}}]{Bitsch15a}
{Bitsch}, B., {Johansen}, A., {Lambrechts}, M., \& {Morbidelli}, A. 2015, \aap,
  575, A28

\bibitem[{{Bitsch} {et~al.}(2018){Bitsch}, {Morbidelli}, {Johansen}, {Lega},
  {Lambrechts}, \& {Crida}}]{Bitsch18}
{Bitsch}, B., {Morbidelli}, A., {Johansen}, A., {et~al.} 2018, \aap, 612, A30

\bibitem[{{Bodenheimer} \& {Pollack}(1986)}]{BodenheimerPollack86}
{Bodenheimer}, P. \& {Pollack}, J.~B. 1986, \icarus, 67, 391

\bibitem[{{Br{\"u}gger} {et~al.}(2018){Br{\"u}gger}, {Alibert}, {Ataiee}, \&
  {Benz}}]{Brugger18}
{Br{\"u}gger}, N., {Alibert}, Y., {Ataiee}, S., \& {Benz}, W. 2018, \aap, 619,
  A174

\bibitem[{{Burn} {et~al.}(2019){Burn}, {Marboeuf}, {Alibert}, \&
  {Benz}}]{Burn19}
{Burn}, R., {Marboeuf}, U., {Alibert}, Y., \& {Benz}, W. 2019, \aap, 629, A64

\bibitem[{{Chambers}(2006)}]{Chambers06}
{Chambers}, J. 2006, \icarus, 180, 496

\bibitem[{{Clarke} {et~al.}(2001){Clarke}, {Gendrin}, \&
  {Sotomayor}}]{Clarke2001}
{Clarke}, C.~J., {Gendrin}, A., \& {Sotomayor}, M. 2001, \mnras, 328, 485

\bibitem[{{Coleman} {et~al.}(2019){Coleman}, {Leleu}, {Alibert}, \&
  {Benz}}]{Coleman19}
{Coleman}, G.~A.~L., {Leleu}, A., {Alibert}, Y., \& {Benz}, W. 2019, \aap, 631,
  A7

\bibitem[{{Coleman} \& {Nelson}(2014)}]{ColemanNelson14}
{Coleman}, G.~A.~L. \& {Nelson}, R.~P. 2014, \mnras, 445, 479

\bibitem[{{Coleman} \& {Nelson}(2016{\natexlab{a}})}]{ColemanNelson16b}
{Coleman}, G.~A.~L. \& {Nelson}, R.~P. 2016{\natexlab{a}}, \mnras, 460, 2779

\bibitem[{{Coleman} \& {Nelson}(2016{\natexlab{b}})}]{ColemanNelson16}
{Coleman}, G.~A.~L. \& {Nelson}, R.~P. 2016{\natexlab{b}}, \mnras, 457, 2480

\bibitem[{{Coleman} {et~al.}(2017){Coleman}, {Papaloizou}, \& {Nelson}}]{CPN17}
{Coleman}, G.~A.~L., {Papaloizou}, J.~C.~B., \& {Nelson}, R.~P. 2017, \mnras,
  470, 3206

\bibitem[{{Crida} {et~al.}(2006){Crida}, {Morbidelli}, \& {Masset}}]{Crida}
{Crida}, A., {Morbidelli}, A., \& {Masset}, F. 2006, \icarus, 181, 587

\bibitem[{{Dr{\k{a}}{\.z}kowska} \& {Alibert}(2017)}]{Drazkowska17}
{Dr{\k{a}}{\.z}kowska}, J. \& {Alibert}, Y. 2017, \aap, 608, A92

\bibitem[{{Fortier} {et~al.}(2013){Fortier}, {Alibert}, {Carron}, {Benz}, \&
  {Dittkrist}}]{Fortier13}
{Fortier}, A., {Alibert}, Y., {Carron}, F., {Benz}, W., \& {Dittkrist}, K.~M.
  2013, \aap, 549, A44

\bibitem[{{Friedrich} {et~al.}(2015){Friedrich}, {Weisberg}, {Ebel}, {Biltz},
  {Corbett}, {Iotzov}, {Khan}, \& {Wolman}}]{Friedrich2015}
{Friedrich}, J.~M., {Weisberg}, M.~K., {Ebel}, D.~S., {et~al.} 2015, Chemie der
  Erde / Geochemistry, 75, 419

\bibitem[{{Haisch} {et~al.}(2001){Haisch}, {Lada}, \& {Lada}}]{Haisch01}
{Haisch}, Karl~E., J., {Lada}, E.~A., \& {Lada}, C.~J. 2001, \apjl, 553, L153

\bibitem[{{Hansen}(2008)}]{Hansen2008}
{Hansen}, B. M.~S. 2008, \apjs, 179, 484

\bibitem[{{Hueso} \& {Guillot}(2005)}]{Hueso05}
{Hueso}, R. \& {Guillot}, T. 2005, \aap, 442, 703

\bibitem[{{Ida} \& {Guillot}(2016)}]{IdaGuillot16}
{Ida}, S. \& {Guillot}, T. 2016, \aap, 596, L3

\bibitem[{{Ida} \& {Lin}(2004)}]{IdaLin2004}
{Ida}, S. \& {Lin}, D.~N.~C. 2004, \apj, 616, 567

\bibitem[{{Inaba} \& {Ikoma}(2003)}]{Inaba}
{Inaba}, S. \& {Ikoma}, M. 2003, \aap, 410, 711

\bibitem[{{Inaba} {et~al.}(2001){Inaba}, {Tanaka}, {Nakazawa}, {Wetherill}, \&
  {Kokubo}}]{Inaba01}
{Inaba}, S., {Tanaka}, H., {Nakazawa}, K., {Wetherill}, G.~W., \& {Kokubo}, E.
  2001, \icarus, 149, 235

\bibitem[{{Jin} {et~al.}(2014){Jin}, {Mordasini}, {Parmentier}, {van Boekel},
  {Henning}, \& {Ji}}]{JinMordasini14}
{Jin}, S., {Mordasini}, C., {Parmentier}, V., {et~al.} 2014, \apj, 795, 65

\bibitem[{{Johansen} \& {Lambrechts}(2017)}]{Johansen17}
{Johansen}, A. \& {Lambrechts}, M. 2017, Annual Review of Earth and Planetary
  Sciences, 45, 359

\bibitem[{{Lambrechts} \& {Johansen}(2012)}]{Lambrechts12}
{Lambrechts}, M. \& {Johansen}, A. 2012, \aap, 544, A32

\bibitem[{{Lambrechts} \& {Johansen}(2014)}]{Lambrechts14}
{Lambrechts}, M. \& {Johansen}, A. 2014, \aap, 572, A107

\bibitem[{{Lenz} {et~al.}(2019){Lenz}, {Klahr}, \& {Birnstiel}}]{Lenz19}
{Lenz}, C.~T., {Klahr}, H., \& {Birnstiel}, T. 2019, \apj, 874, 36

\bibitem[{{Lin} \& {Papaloizou}(1986)}]{LinPapaloizou86}
{Lin}, D.~N.~C. \& {Papaloizou}, J. 1986, \apj, 309, 846

\bibitem[{{Lodders}(2003)}]{Lodders2003}
{Lodders}, K. 2003, \apj, 591, 1220

\bibitem[{{Lynden-Bell} \& {Pringle}(1974)}]{Lynden-BellPringle1974}
{Lynden-Bell}, D. \& {Pringle}, J.~E. 1974, \mnras, 168, 603

\bibitem[{{Mamajek}(2009)}]{Mamajek09}
{Mamajek}, E.~E. 2009, in American Institute of Physics Conference Series, Vol.
  1158, American Institute of Physics Conference Series, ed. T.~{Usuda},
  M.~{Tamura}, \& M.~{Ishii}, 3--10

\bibitem[{{Matsuyama} {et~al.}(2003){Matsuyama}, {Johnstone}, \&
  {Hartmann}}]{Matsuyama03}
{Matsuyama}, I., {Johnstone}, D., \& {Hartmann}, L. 2003, \apj, 582, 893

\bibitem[{{Morbidelli} {et~al.}(2015){Morbidelli}, {Lambrechts}, {Jacobson}, \&
  {Bitsch}}]{Morbi15}
{Morbidelli}, A., {Lambrechts}, M., {Jacobson}, S., \& {Bitsch}, B. 2015,
  \icarus, 258, 418

\bibitem[{{Mordasini} {et~al.}(2009){Mordasini}, {Alibert}, \&
  {Benz}}]{Mords09}
{Mordasini}, C., {Alibert}, Y., \& {Benz}, W. 2009, \aap, 501, 1139

\bibitem[{{Mordasini} {et~al.}(2012{\natexlab{a}}){Mordasini}, {Alibert},
  {Georgy}, {Dittkrist}, {Klahr}, \& {Henning}}]{Mord12b}
{Mordasini}, C., {Alibert}, Y., {Georgy}, C., {et~al.} 2012{\natexlab{a}},
  \aap, 547, A112

\bibitem[{{Mordasini} {et~al.}(2012{\natexlab{b}}){Mordasini}, {Alibert},
  {Klahr}, \& {Henning}}]{Mord12a}
{Mordasini}, C., {Alibert}, Y., {Klahr}, H., \& {Henning}, T.
  2012{\natexlab{b}}, \aap, 547, A111

\bibitem[{{Mordasini} {et~al.}(2014){Mordasini}, {Klahr}, {Alibert}, {Miller},
  \& {Henning}}]{Mordasini14}
{Mordasini}, C., {Klahr}, H., {Alibert}, Y., {Miller}, N., \& {Henning}, T.
  2014, \aap, 566, A141

\bibitem[{{Mordasini} {et~al.}(2015){Mordasini}, {Molli{\`e}re}, {Dittkrist},
  {Jin}, \& {Alibert}}]{Mordasini15}
{Mordasini}, C., {Molli{\`e}re}, P., {Dittkrist}, K.-M., {Jin}, S., \&
  {Alibert}, Y. 2015, International Journal of Astrobiology, 14, 201

\bibitem[{{Nakamoto} \& {Nakagawa}(1994)}]{Nakamoto94}
{Nakamoto}, T. \& {Nakagawa}, Y. 1994, \apj, 421, 640

\bibitem[{{Ohtsuki}(1999)}]{Ohtsuki99}
{Ohtsuki}, K. 1999, \icarus, 137, 152

\bibitem[{{Ormel} \& {Klahr}(2010)}]{OrmelKlahr2010}
{Ormel}, C.~W. \& {Klahr}, H.~H. 2010, \aap, 520, A43

\bibitem[{{Paardekooper} {et~al.}(2010){Paardekooper}, {Baruteau}, {Crida}, \&
  {Kley}}]{pdk10}
{Paardekooper}, S.-J., {Baruteau}, C., {Crida}, A., \& {Kley}, W. 2010, \mnras,
  401, 1950

\bibitem[{{Paardekooper} {et~al.}(2011){Paardekooper}, {Baruteau}, \&
  {Kley}}]{pdk11}
{Paardekooper}, S.-J., {Baruteau}, C., \& {Kley}, W. 2011, \mnras, 410, 293

\bibitem[{{Paardekooper} \& {Mellema}(2006)}]{PaardekooperMellema06}
{Paardekooper}, S.-J. \& {Mellema}, G. 2006, \aap, 459, L17

\bibitem[{{Pollack} {et~al.}(1996){Pollack}, {Hubickyj}, {Bodenheimer},
  {Lissauer}, {Podolak}, \& {Greenzweig}}]{Pollack}
{Pollack}, J.~B., {Hubickyj}, O., {Bodenheimer}, P., {et~al.} 1996, \icarus,
  124, 62

\bibitem[{{Rafikov}(2004)}]{Rafikov04}
{Rafikov}, R.~R. 2004, \aj, 128, 1348

\bibitem[{{Ruden} \& {Pollack}(1991)}]{Ruden91}
{Ruden}, S.~P. \& {Pollack}, J.~B. 1991, \apj, 375, 740

\bibitem[{{Santos} {et~al.}(2005){Santos}, {Israelian}, {Mayor}, {Bento},
  {Almeida}, {Sousa}, \& {Ecuvillon}}]{Santos05}
{Santos}, N.~C., {Israelian}, G., {Mayor}, M., {et~al.} 2005, \aap, 437, 1127

\bibitem[{{Saumon} {et~al.}(1995){Saumon}, {Chabrier}, \& {van
  Horn}}]{Saumon95}
{Saumon}, D., {Chabrier}, G., \& {van Horn}, H.~M. 1995, \apjs, 99, 713

\bibitem[{{Schoonenberg} {et~al.}(2019){Schoonenberg}, {Liu}, {Ormel}, \&
  {Dorn}}]{Schoonenberg19}
{Schoonenberg}, D., {Liu}, B., {Ormel}, C.~W., \& {Dorn}, C. 2019, \aap, 627,
  A149

\bibitem[{{Scott}(1992)}]{Scott92}
{Scott}, D.~W. 1992, {Multivariate Density Estimation}

\bibitem[{{Shakura} \& {Sunyaev}(1973)}]{Shak}
{Shakura}, N.~I. \& {Sunyaev}, R.~A. 1973, \aap, 24, 337

\bibitem[{{Shibaike} {et~al.}(2019){Shibaike}, {Ormel}, {Ida}, {Okuzumi}, \&
  {Sasaki}}]{Shibaike19}
{Shibaike}, Y., {Ormel}, C.~W., {Ida}, S., {Okuzumi}, S., \& {Sasaki}, T. 2019,
  \apj, 885, 79

\bibitem[{{Tychoniec} {et~al.}(2018){Tychoniec}, {Tobin}, {Karska}, {Chand
  ler}, {Dunham}, {Harris}, {Kratter}, {Li}, {Looney}, {Melis}, {P{\'e}rez},
  {Sadavoy}, {Segura-Cox}, \& {van Dishoeck}}]{Tychoniec18}
{Tychoniec}, {\L}., {Tobin}, J.~J., {Karska}, A., {et~al.} 2018, \apjs, 238, 19

\end{thebibliography}
